\documentclass[final,authoryear,5p,times]{elsarticle}

\usepackage{amssymb}
\usepackage{amsmath}
\usepackage{hyperref}
\usepackage{booktabs}
\usepackage{multirow}
\usepackage{caption}  
\usepackage{microtype}
\usepackage{cleveref}
\crefname{appendix}{}{}

\renewcommand{\b}{\mathbf}
\DeclareMathOperator*{\argmax}{arg\,max}
\DeclareMathOperator*{\argmin}{arg\,min}

\usepackage{bm}

\usepackage{color}

\journal{Astronomy and Computing}

\begin{document}

\begin{frontmatter}



\title{Detecting Dispersed Radio Transients in Real Time Using Convolutional Neural Networks}


\author[ai4s, api]{David Ruhe\corref{cor}}
\ead{d.ruhe@uva.nl}
\author[api]{Mark Kuiack}
\author[api]{Antonia Rowlinson}
\author[api]{Ralph Wijers}
\author[ai4s]{Patrick Forr\'e}

\fntext[api]{Anton Pannekoek Institute, University of Amsterdam}
\fntext[ai4s]{AI4Science (AMLab), Informatics Institute, University of Amsterdam}

\cortext[cor]{Corresponding author at: Anton Pannekoek Institute, University of Amsterdam, Amsterdam, 1098XH, The Netherlands}


\begin{abstract}
We present a methodology for automated real-time analysis of a radio image data stream with the goal to find transient sources.
Contrary to previous works, the transients we are interested in occur on a time-scale where dispersion starts to play a role, so we must search a higher-dimensional data space and yet work fast enough to keep
up with the data stream in real time.
The approach consists of five main steps: quality control, source detection, association, flux measurement, and physical parameter inference.
We present parallelized methods based on convolutions and filters that can be accelerated on a GPU, allowing the pipeline to run in real-time.
In the parameter inference step, we apply a convolutional neural network to dynamic spectra that were obtained from the preceding steps. 
It infers physical parameters, among which the dispersion measure of the transient candidate.
Based on critical values of these parameters, an alert can be sent out and data will be saved for further investigation.
Experimentally, the pipeline is applied to simulated data and images from AARTFAAC (Amsterdam Astron Radio Transients Facility And Analysis Centre), a transients facility based on the Low-Frequency Array (LOFAR).
Results on simulated data show the efficacy of the pipeline, and from real data it discovered dispersed pulses.
The current work targets transients on time scales that are longer than the fast transients of beam-formed search, but shorter than slow transients in which dispersion matters less.
This fills a methodological gap that is relevant for the upcoming Square-Kilometer Array (SKA).
Additionally, since real-time analysis can be performed, only data with promising detections can be saved to disk, providing a solution to the big-data problem that modern astronomy is dealing with.

\end{abstract}


\end{frontmatter}


\section{Introduction}
    
\begin{figure*}
    \centering
    \includegraphics[width=0.4\linewidth]{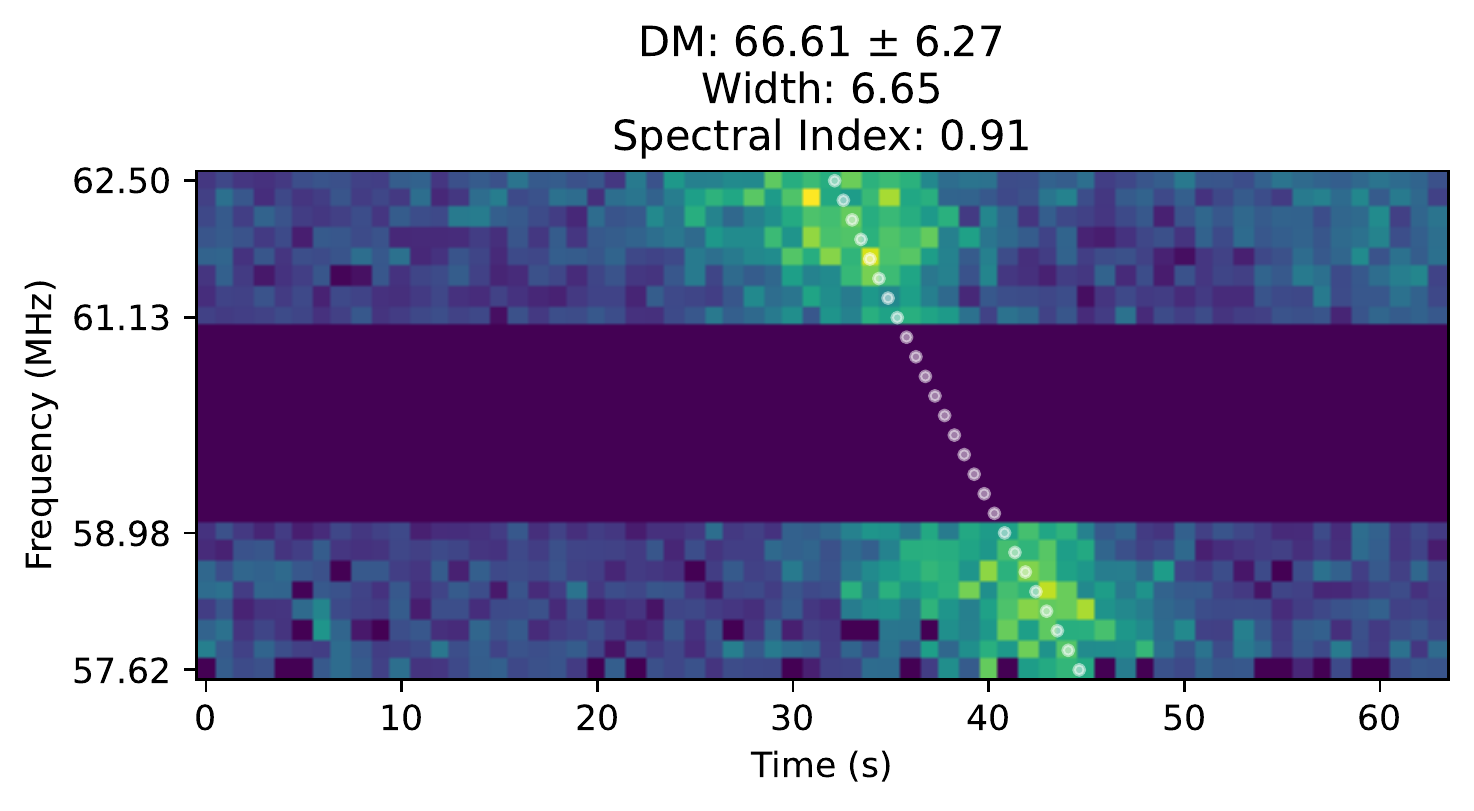}
    \includegraphics[width=0.4\linewidth]{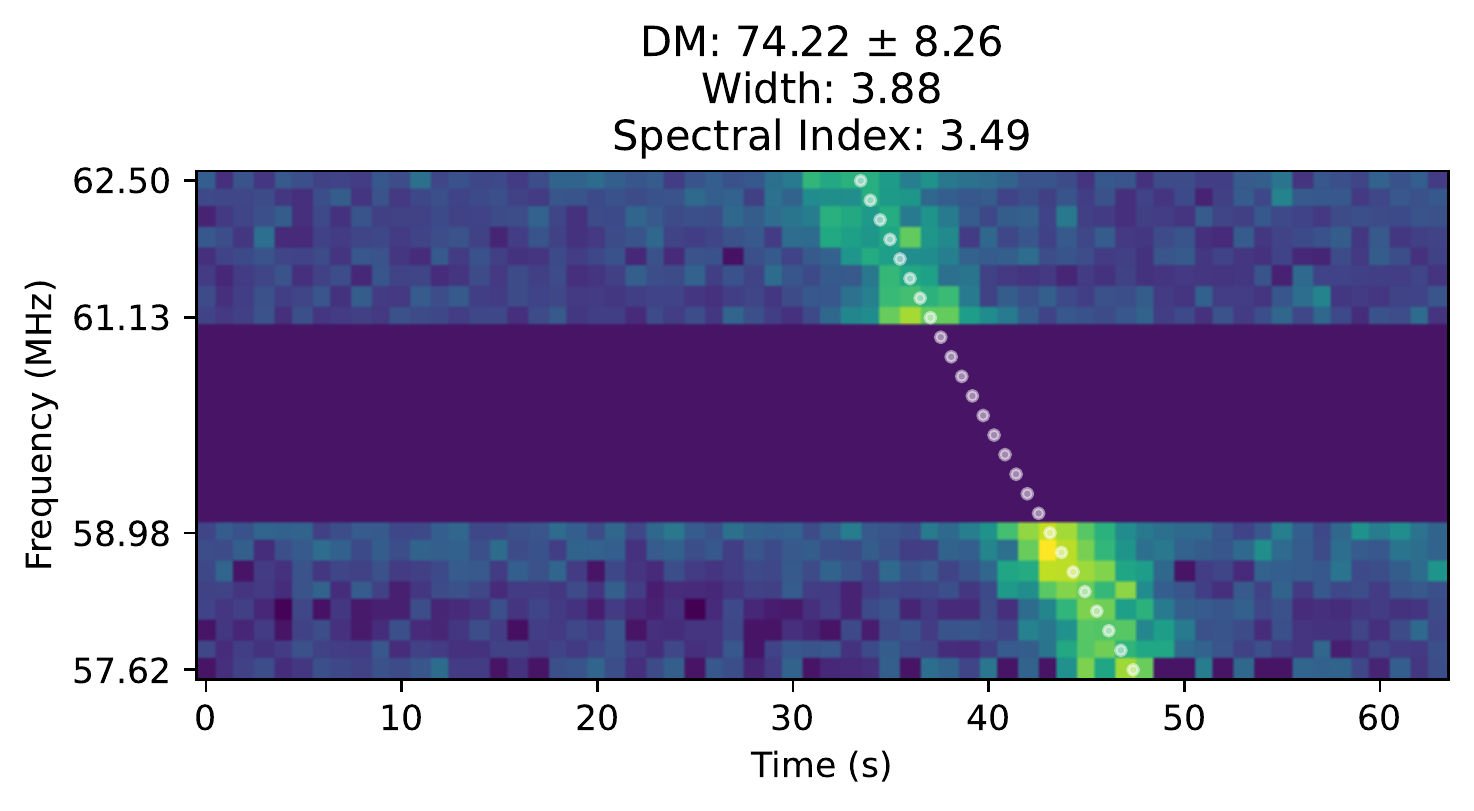} \\
    \includegraphics[width=0.4\linewidth]{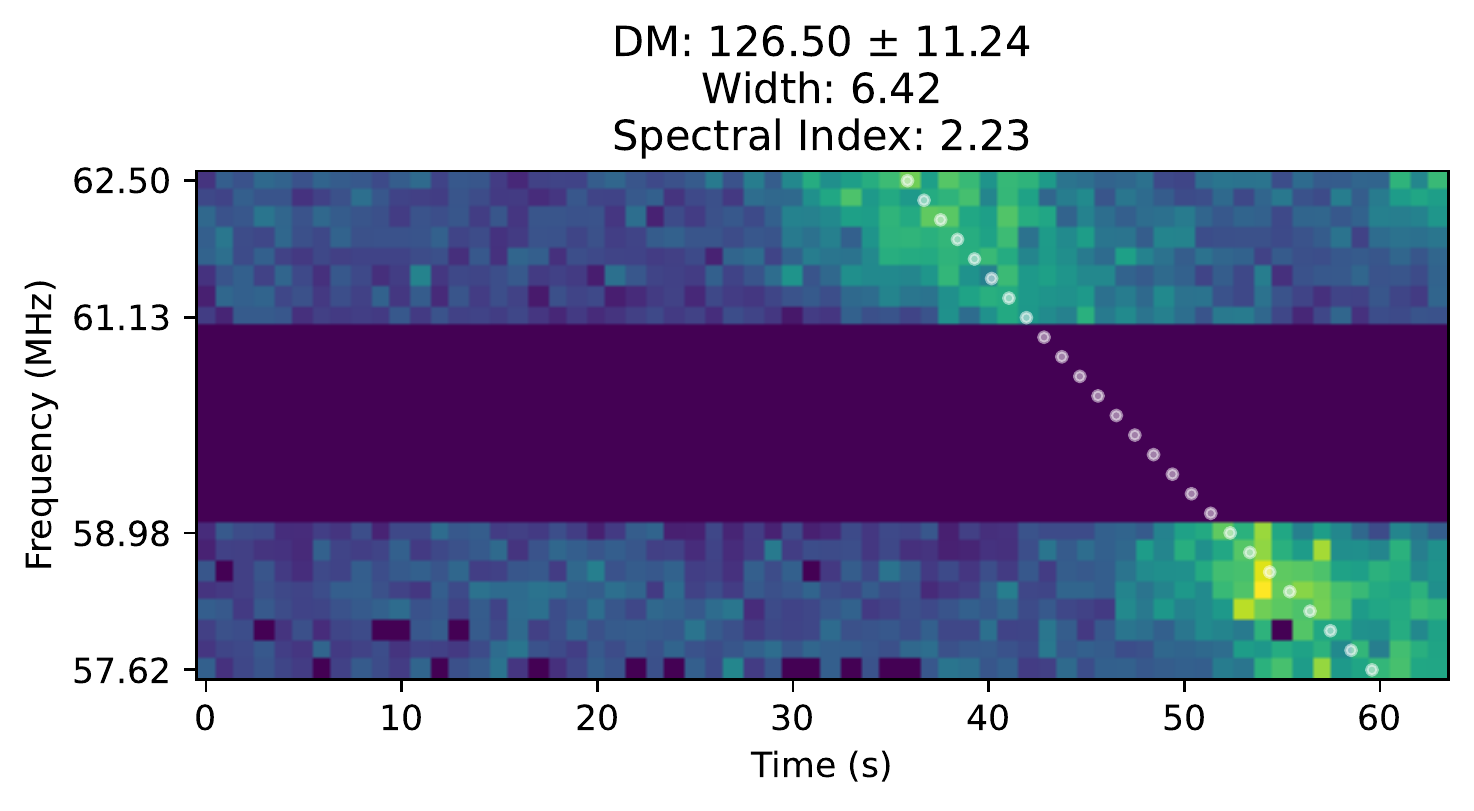}
    \includegraphics[width=0.4\linewidth]{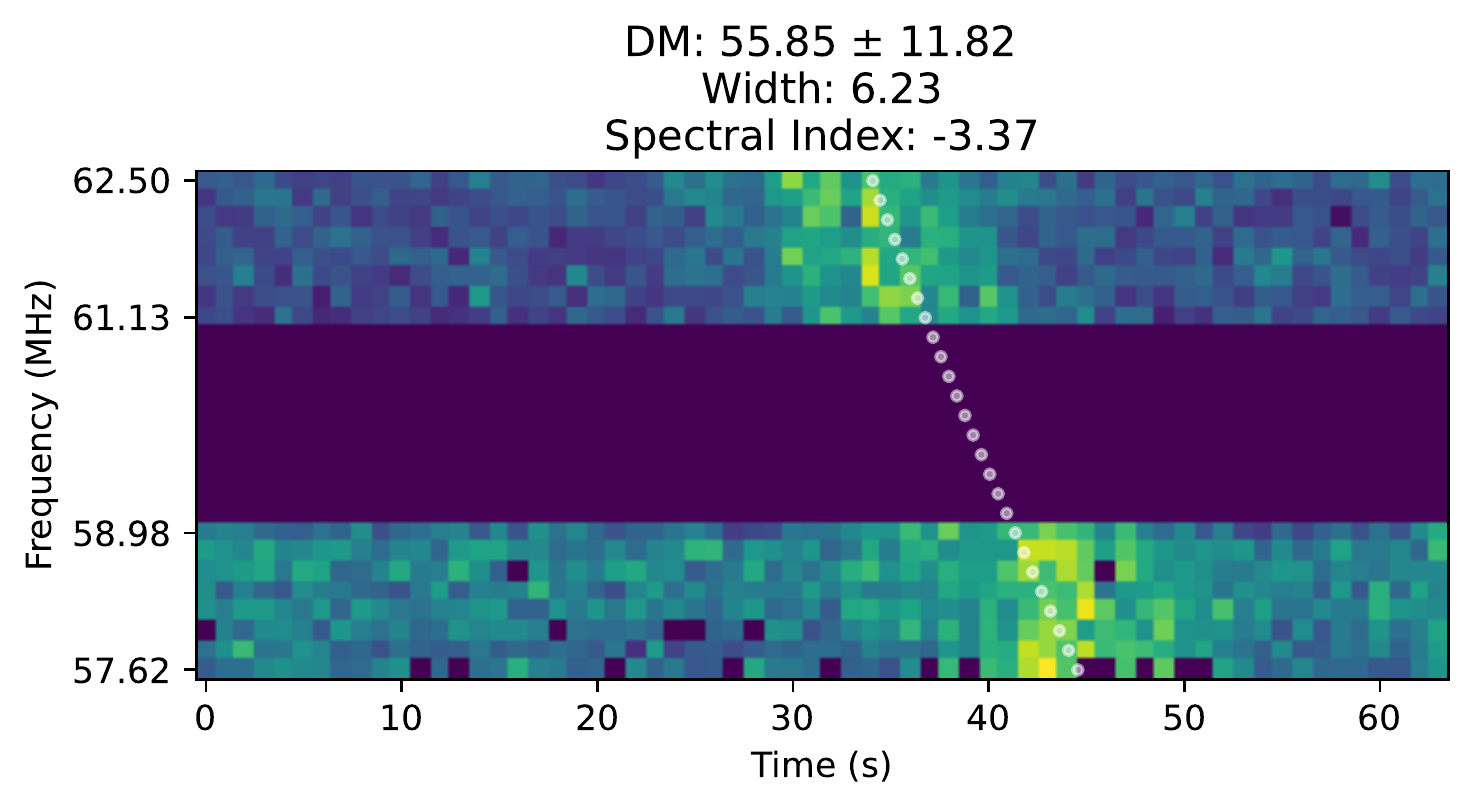}
    \caption{
    Examples of dispersed signals that were found after applying the pipeline. 
    The $y$-axis denotes frequency, where the the time in seconds is shown on the $x$-axis. 
    The instrument (AARTFAAC) observed at two separate ranges of consecutive bands, resulting in the empty central gap.
    The parameters $\hat{\bm \theta}$ inferred by the convolutional neural network are provided above the respective samples. 
    Additionally, we plot the dispersion sweep according to the inferred dispersion measure $\mathrm{DM}$.
    The upper-right example is the proposed transient candidate by \cite{kuiack2020aartfaac}.
    The other three are yet to be confirmed new transient candidates.
    }
    \label{fig:examples}
\end{figure*}
The advent of instruments that have large fields of view in relatively unexplored frequency domains has boosted the interest for blind transient searches \citep{shin2009detecting, bannister201122, bower2011allen, thyagarajan2011variable, hoffman2012variability,  franzen2014deep, bell2014survey, ivezic2019lsst,  kuiack2020aartfaac, villar2021deep}. 
The low radio regime now also has entered all-sky astronomy with transients being one of the key objectives \citep{bell2011automated, taylor2012first, murphy2013vast, pintaldi2021scalable}.

Much work has been done on detecting fast radio transients \citep{cordes2003searches, lorimer2013detectability, coenen2014lofar, amiri2018chime} that occur on millisecond time scales.
This is usually realized using beamforming \citep{lorimer2012handbook}, having high time and frequency resolution at the cost of poor spatial resolution.
The arguably foremost example of such fast transients is the fast radio burst \citep{petroff2019fast}.
One of the key features of fast radio transients is the dispersion of the observed emission in time and frequency, in which emission at lower radio frequencies arrive later in time than the emission at higher radio frequencies \citep{1993ApJ...411..674T}. 
Each source has a characteristic Dispersion Measure (DM); more distant sources have a higher DM value corresponding to a larger delay between receiving the high and low frequency emission.

Alternatively, researchers have analyzed slow transients, occurring on time scales where dispersion plays a minor role \citep{williams2012asgard, murphy2013vast, chen2013long, rowlinson2016limits, murphy2017search, law2018discovery}.
These are usually detected in image data.

In between these two extremes, we enter a domain where we investigate images of a large field of view at a relatively high time resolution.
The high time-resolution means that the transients (of intermediate length, lasting seconds to minutes) can be significantly dispersed, especially at low radio frequencies.
Consequently, we also need sufficient frequency resolution.
Altogether, analysis is to be performed in the spatial, temporal, and frequency domains.
This is a higher-dimensional search space than the previously introduced methods.
On the one hand beamformed search only considers the time and frequency domains.
On the other hand, slow transient search considers mostly the spatial and time domains.
\cite{kuiack2020long, kuiack2020apparent} show such intermediate length transient detections in this type of data. 
These were achieved in offline analysis, causing a long latency between a transient's occurrence and its discovery. 
If we can find them in real time, we enable follow-up studies before they have faded. 
This greatly enhances the scientific return, making a strong case for further development of structured search methods.
We note, of course, that imaging data usually have much lower time and frequency resolution
than beam-formed data, so while we will search a higher-dimensional data space, we do not necessarily 
search a larger data volume.

At every time step, we obtain an image cube containing a frequency dimension and two spatial dimensions.
Analysis pipelines for these have been proposed before by e.g. \cite{swinbank2015lofar}.
However, these works are not yet scalable enough to perform blind transient searches in the enormous volumes of data.
Additionally, there have been targeted searches for specific types of highly dispersed sources in high time and frequency resolution data obtained using the Murchison Widefield Array \citep[MWA; ][]{tingay2013murchison}. 
\cite{2015AJ....150..199T} piloted a search for dispersed fast radio bursts in high time and frequency resolution imaging data covering a 400 square degree field of view obtained using the MWA. 
However, processing just 2 hours of data required 3 days on a single processing core, making this computationally inefficient and far from attaining real-time analysis. 
Searches using the MWA have also focused on highly dispersed sources expected to be detected in just a single pixel in the radio image, enabling a significant speed up in processing time but at the sacrifice of not searching the full field of view \citep[e.g. ][]{2021PASA...38...26A}. 
Thus, there is a need for a computationally efficient method to search for dispersed radio transients in wide field of view imaging observations.
We thereby can blindly search for the brightest and rarest transients, filtering only the most useful information and alerting the multi-wavelength transient community when required.
This means that the enormous number of spurious candidates from radio frequency interference (RFI), scintillating sources, and random noise have to be filtered automatically.


\begin{figure*}
    \centering
    \includegraphics[width=0.65\textwidth]{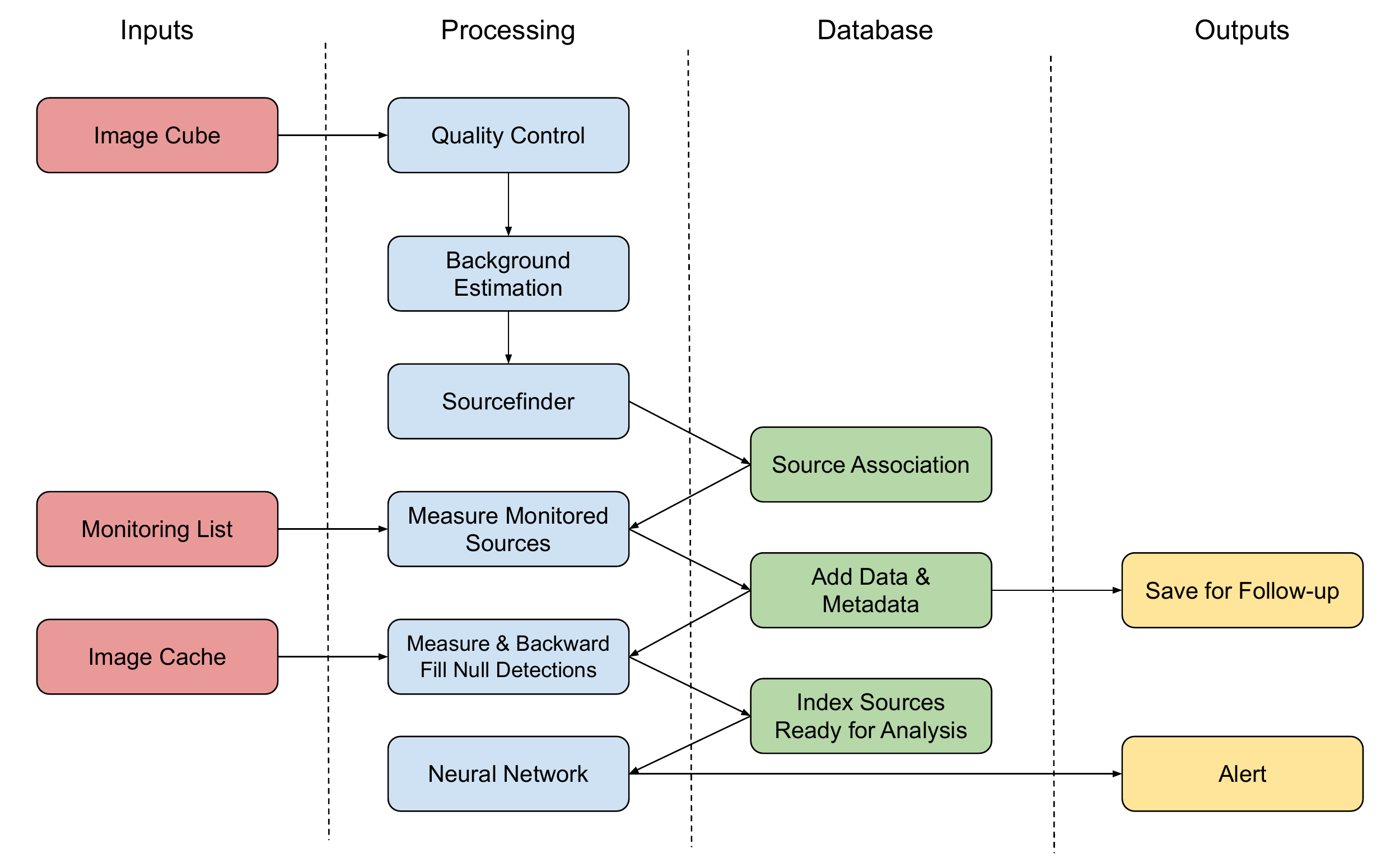}
    \caption{Overview of all the steps in the proposed pipeline.}
    \label{fig:pipeline}
\end{figure*}
    
As an answer to the aforementioned, we develop a pipeline that considers the spatial, time, and frequency domain simultaneously and can detect dispersed transients of intermediate length.
It scales to real-time analysis by its ability to run on GPUs.
Five sequential steps are performed: quality control, source detection, source association, flux measurement, and physical parameter inference.
Motivated before, we want to keep a high-frequency resolution and therefore do not integrate the frequency bands.
This allows for new processing approaches.
We can control the quality of the images (step 1) by comparing them to the other bands.
Next, it allows for source detection (step 2) in sub-bands independently.
Furthermore, we measure the source flux (step 4) in separate bands.
We propose methods for source detection and measurement using convolutions and filters, which are easily parallelizable.
Since we perform the processing on the sub-bands individually, we can directly construct dynamic spectra from which we infer physical parameters of a potential transient candidate (step 5).
We apply a convolutional neural network machine learning approach to do so.
Particularly the DM is of interest in separating spurious from real transients.
By doing so, we rethink the challenge of detecting dispersed transients by using dispersion directly to discard spurious candidates.

We test our approaches on simulated and real data from AARTFAAC \citep[Amsterdam Astron Radio Transients Facility And Analysis Centre; ][]{prasad2016aartfaac}, a real-time transients facility based on the Low-Frequency Array \citep[LOFAR; ][]{van2013lofar}.
However, our methods could also be implemented by MWA, Long Wavelength Array Station 1 \citep[LWA1; ][]{obenberger2014limits}, Owens Valley Radio Observatory Long Wavelength Array \citep[OVRO-LWA;][]{anderson2019new} LOFAR, and the future Square-Kilometer Array \citep[SKA; ][]{carilli2004science}.
The results show that the detection methods can reliably find transient candidates, and the neural network discriminates spurious candidates from promising ones using reliable uncertainty bounds.
As shown in \cref{fig:examples}, interesting bursts are uncovered, among which the candidate proposed by \cite{kuiack2020aartfaac}.
Moreover, the pipeline can perform these steps in real time, allowing for online selection of data to be saved to disk for follow-up investigation.

Our scientific contributions can be summarized as follows:
\begin{itemize}
    \item We propose an end-to-end GPU-accelerated pipeline that can take streaming multi-frequency image data and output alerts in real time. It contains source detection, tracking (i.e., association) and analysis\footnote{Code is publicly available at \url{https://ascl.net/2103.015} \citep{2021ascl.soft03015R}}.
    Our method is the first that considers the spatial, time, and frequency domains of the incoming data simultaneously. This allows for transient hunting on intermediate time scales.
    \item We propose source detection and measurement methods based on convolutions and filters.
    \item A neural-network-based analysis approach, in which physical parameters are inferred directly from dynamic spectra.
\end{itemize}

The paper is organized as follows. 
In \cref{sec:methodology} we describe the end-to-end pipeline and its methodologies.
In \cref{sec:results} we report results of experiments that were done to test the pipeline and some preliminary data products extracted from application to observations. 
Finally, we conclude in \cref{sec:conclusion} and foresee some interesting directions for further research and development.

\section{Methodology}
\label{sec:methodology}
An overview of our transients pipeline is given in \cref{fig:pipeline}.
In this section, we discuss some of the approaches.
Since it is somewhat specific to our instrument, the methods for quality control of the input data can be found in \cref{sec:supp_quality_control}.

\subsection{Source Detection}

\label{sec:source_detection}
\begin{figure}
    \centering
    \includegraphics[width=0.8\linewidth]{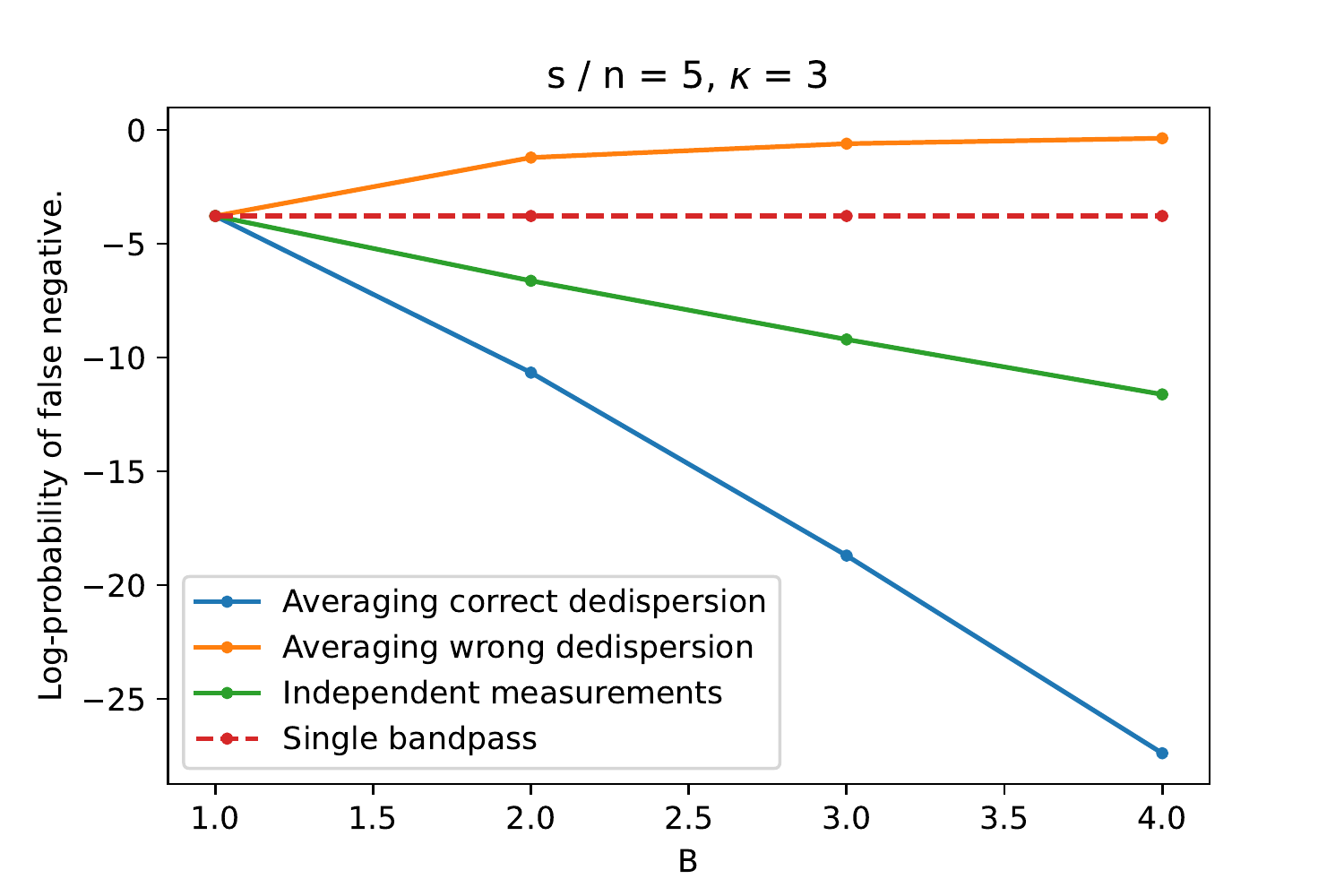} \\
    \caption{Taken from \cref{app:statistics}. False-negative rate as a function of the number of bandpasses $B$ (lower is better). As a reference, we show the log-probability for a single pass. When not or incorrectly adjusting for dispersion, the probability of a false negative goes up. For the given signal-to-noise-ratio $s/n$ and $\kappa$, the optimal case (but practically unattainable) is averaging after correctly adjusting for dispersion. Independently searching also gives significant false negative error improvement as $B$ grows.}
    \label{fig:type_2_error_example}
\end{figure}

The input at time $t$ to our pipeline is an image cube $\b X_t \in \mathbb{R}^{B\times D \times D}$, where $B$ is the number of channels and $D$ the image size.
We analyze images at multiple bandpasses in parallel.
This is done since astronomical transients are expected to be dispersed over these bandpasses.
By doing so, the probability of a false negative (FN) goes down with $B$ as
\begin{equation}
    \mathrm{Pr}\left(\mathrm{FN}\right) = \left[\Phi\left(\kappa - s/n \right)\right]^B
    \label{eq:pr_fn}
\end{equation}
where $\Phi$ is the standard cumulative density function, $\kappa$ user-defined and $s/n$ the signal-to-noise ratio of the source.
When perfectly correcting for dispersion, one usually gets a factor $\sqrt B$ (times $s/n$) decrease.
However, applying brute force coherent dedispersion before searching is practically unattainable in real time in image space for all sources.
Still, \cref{eq:pr_fn} is always better than not or even wrongly using the bandpasses.
Aggregating the bandpasses \citep[as in e.g.][]{anderson2019new} without correcting for dispersion would increase the probability of a false negative with $\sqrt B$.
The above is illustrated in \cref{fig:type_2_error_example}.
Derivations and further discussion are presented in \cref{app:statistics}.

\subsubsection{Peak Detection}
Consider a single image $X \in \mathbb{R}^{D \times D}$ (time-index omitted) from $\b X_t$.
Previous approaches \citep[e.g. ][]{swinbank2015lofar, spreeuw2018pyse} divide the image to be analyzed into a grid. 
In every grid cell, sigma-clipping is performed and thus local statistics are used for the detection threshold.
We replace this approach with a method that combines sigma-clipping with convolutions.
When dividing the image into a grid, an isolated faint source might not be detected when it is at the edge of a grid patch that is crowded with bright sources (or radio-frequency interference). 
The bright objects first have to be clipped away for the source to be found.
This is usually countered (naively) by interpolating the grid.
Alternatively, one might want to sigma-clip using overlapping grid cells, lowering the risk that a faint source at an unfortunate position might not be picked up on.
However, taken to the extreme case this becomes a convolution where we use grid statistics at every pixel coordinate.
By the convolution theorem, this can effectively be implemented using Fourier transforms and accelerated on GPUs.

Let 
\begin{equation}
G(n, m, s) := \frac1Z \exp \left[-\frac1{2 s^2} \left(n^2 + m^2\right)\right]
\end{equation} 
be a Gaussian kernel with 
$Z := \sum_{i=n-k}^{n+k} \sum_{j=m-k}^{m + k} \exp\left[-\frac{1}{2s^2}(i^2 + j^2)\right]$
its normalizing constant.
Alternatively, one could also go for a (circular) uniform kernel.
$s$ is an important parameter and should be set such that the local background statistics are sufficiently included, without incorporating too much source flux.
Heuristics for setting it are e.g. widening the Gaussian such that its full width at half maximum (FWHM) includes most primary side-lobes, or running an optimization scheme such that it retrieves as many as possible known sources from a reference catalog.
The following is run for multiple iterations.
At each iteration, we have
\begin{equation}
    C := G * X
\end{equation}
and 
\begin{equation}
    S := G * (X - C)^2
\end{equation}
where the square is applied element-wise.
$C_{nm}$ and $S_{nm}$ are center and spread estimates of the noise computed at every location $nm$.
We clip values in the following manner:
\begin{equation}
    y_{nm} := \mathbb{I}\left(x_{nm} > c_{nm} + \kappa \sqrt{s_{nm}}\right)
\end{equation}
with $\kappa$ as a parameter that specifies how much signal is required for a detection to be made.
This procedure is repeated until no new $y_{nm}$ is discovered or for a pre-specified number of iterations.
Since we have accurate noise estimates at every $nm$, not many iterations are needed to find most of the true positives (\cref{sec:supp_sigmaclip_iterations}).
The result of the iterative sigma-clipping done is a binary image $Y \in \{0, 1 \}^{D \times D}$  indicating the source locations.

\subsubsection{Peak Localization}
The sources in $Y$ usually are extended (i.e., they span multiple connected locations).
In order to isolate them to a single coordinate, we apply a maximum filter to $X$ and compute $Q\in \{0, 1 \}^{D \times D}$ indicating if the pixels are the local maximum:
\begin{align}
    q_{nm} := \mathbb{I}\big( &x_{nm} = \max\{x_{ij} \mid i, j \nonumber \\ &\in  \{ n - k, \dots, n + k \} \times \{ m - k, \dots,    m + k \}\} \big)
\end{align}
with $k=3$.
Then, we obtain the peak locations $P \in \{0, 1\}^{D \times D}$ by the following boolean operation:
\begin{equation}
    p_{nm} := q_{nm} \land y_{nm}.
\end{equation}
Intuitively, we only include pixels that are both the local maximum and above the detection threshold, giving us the exact location of the peaks.
The advantage of doing so (next to parallelization) is that this automatically deblends detected sources. I.e., it separates an ``island'' of flux that exceeds the local noise level into distinct sources.
Obtaining $P$ for all $B$ bandpasses independently gives us a cube $\b P \in \{0, 1\}^{B \times D \times D}$ with source locations.

This concludes the source detection \& localization pipeline.
We have presented new methods based on convolutions and kernels.
They are elegant in that they do not require a discontinuous grid to be placed over the image.
Additionally, they are fully parallel and accelerated on GPUs.
That is, all operations are performed concurrently on separate spatial or channel locations.

\subsection{Source Association \& Flux Measurement}
\label{sec:assoc}
At time-step $t$ after we obtain a list of sources from the source detection pipeline, we filter the duplicates that we obtain from measuring at different frequencies simultaneously.
Next, we match the catalog's previous time-step $t-1$ sources based on the 2D distance (in degrees).
Sources are matched if they are within a pre-specified association distance limit (e.g., 1 degrees).
After initial detection, we keep taking measurements (i.e., monitor the source) until it has not been detected for at least a pre-specified number of time-steps.
This is to obtain data products that are not too sparse for analysis.
We concatenate the detected sources with the sources that are to be monitored.
Next, we take measurements of the fluxes of detected and monitored sources at all frequencies.
This is done by taking the maximum pixel value within a box around the source peak.
In the interest of time, we are not fitting and integrating Gaussians to the detected sources.
We add the measured peak flux to the database for every monitored or newly detected source.
Finally, we backward fill the catalog with ``null detections'' (newly detected sources).
That is, if some potential transient is detected in time-step $t$, we also want to include the ``build-up'' into the data products.
Thus, we add a pre-specified number of timesteps to the catalog and measure from cached images.

\subsection{Neural Network-based Parameter Inference}
\begin{figure*}
    \centering
    \includegraphics[width=\linewidth]{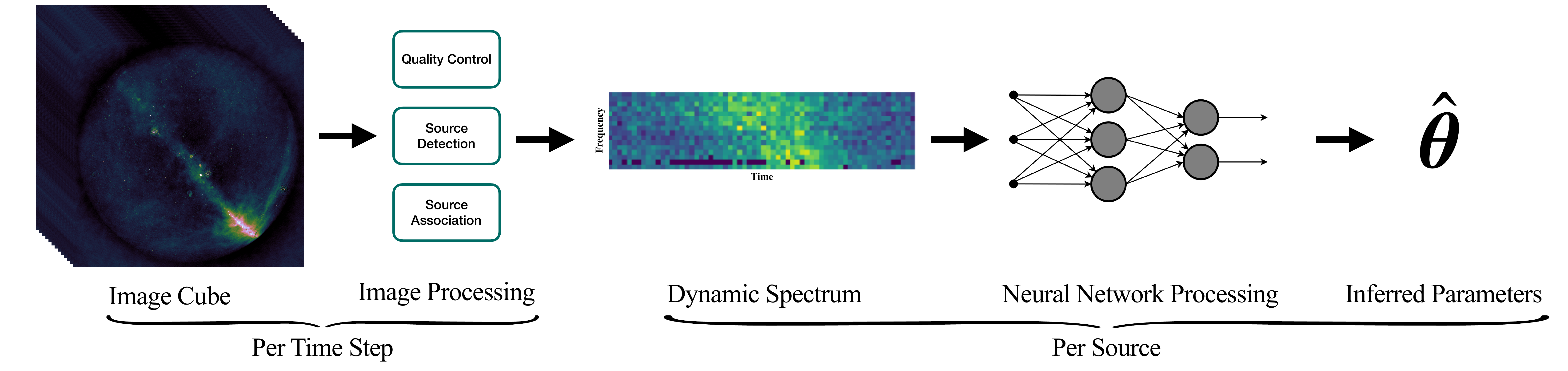} \\
    \caption{Illustrative example of the pipeline. Every time-step an image cube is processed. Collecting the fluxes over time at different bandpasses yields dynamic spectra for all the detected sources. These are processed by a convolutional neural network to infer physical parameters. Spurious candidates can easily be filtered based on their values.}
\end{figure*}
\label{sec:nn}
Our quality control and source detection pipelines already filter out spurious candidates based on (local) noise statistics.
However, in radio astronomy there exist many noise modes (e.g., RFI, satellites, airplanes) that will slip through as candidate transients.
These signals are not dispersed due to their nearness.
This is in contrast with astronomical transients, giving us a key feature to filter them.
This was originally seen by \cite{kuiack2020aartfaac}, who heuristically searched for dispersed candidates.
We take a more systematic approach and resort to machine learning.
Neural networks in particular can process many data instances in parallel, especially on GPUs \citep{lecun2015deep}.
In this subsection, we elaborate on how we use inferred physical parameters to filter the spurious transient candidates.
Afterward, we give an introduction to deep learning and convolutional neural networks \citep{lecun1995convolutional}.
Finally, we explain how these are used to obtain physical parameters.

\subsubsection{Filtering Candidates Based On Physical Parameters}
An astrophysical burst can be described using a range of physical parameters.
For radio transients, these are among others overall shape, integrated and peak flux density, pulse width, dispersion measure, spectral information (such as index and bandwidth), and scattering.
By obtaining these, we can quickly apply a filter and narrow follow-up investigation to only the most interesting cases.
For example, any noise progenitor that is relatively close (e.g., atmospheric or human-made) will not be dispersed since the integrated electron density along the path is too small.
If we can quickly obtain the dispersion measure we can exclude low-DM bursts.
Next, we give an introduction to deep learning and how it is used to infer these parameters of interest.

\subsubsection{Deep Learning}
We briefly discuss the set-up of the neural network 
For a more thorough explanation of how the neural network can be used on frequency-time plots, we refer the reader to e.g. \cite{connor2018applying}.
In its essence, a neural network is a regression preceded by a series of nonlinear transformations.
The idea is that the network learns a mapping from input space to another space (``hidden state'') from which it completes a task (e.g., classification or regression).
The projection into the hidden space allows a neural network to learn a representation.
This representation reflects important features that were computed from the input data.
Let $\b x \in \mathbb{R}^d$ be an arbitrary data-point, $\b w \in \mathbb{R}^d$ a vector of weights, $b \in \mathbb{R}$ a bias value.
A hidden state is computed using a linear combination of the input with a set of weights, and a nonlinearity (referred to as ``activation function'') $\phi(\cdot)$:
\begin{equation}
    h: = \phi \left (\b x^\top \b w + b \right ).
\end{equation}
\ In this work we use Leaky ReLU functions: 
\begin{equation}
\phi(h) := \begin{cases} h \text{ for } h \geq 0 \\ \alpha h \text{ for } h < 0 \end{cases}
\end{equation}
with $\alpha$ a small positive value.
Concatenating multiple such layers with a final task-specific layer forms a neural network.
Since we are dealing with a regression problem here (inferring parameters from the dynamic spectra) our final layer is simply another linear combination but without an activation function.

Convolutional neural networks \citep{lecun1995convolutional} are a type of neural network where the weight vector comes in the form of a filter (or kernel) with which the input image is convolved.
This weight filter is then learned, extracting relevant features from the image.
This can, like a traditional feed-forward network, be optimized with regular gradient descent and backpropagation \citep{lecun2015deep}.
The fact that only kernels (typically multiple per layer) are learned, makes this a lightweight network that is particularly effective for image processing.
Applying the kernels in a convolution also has the advantage that the network output is translationally invariant.
This is an important property since we apply the network to dynamic spectrums in which the burst can occur at multiple spatial locations.


\subsubsection{Parameter Inference}
\label{sec:parameter_inference}
Let $\bm \theta \in \mathbb{R}^{d^\prime}$ be a parameter vector associated with a data-point $\b x$.
In our case, $\b x$ is a dynamic spectrum that we obtained from the image processing steps (see \cref{fig:pipeline}).
We are interested in $p(\bm \theta \mid \b x)$.
Since $\bm \theta$ is real-valued, a natural choice is to model $p(\bm \theta \mid \b x)$ with a Gaussian.
Let 
\begin{equation}
\begin{bmatrix}\hat{\bm \mu} \\ \hat{\b L} \end{bmatrix} := \b g\left(\b x, \b W\right)\end{equation}
be the output of the neural network $\b g(\b x, \b W)$, where $\b W$ is a concatenation of the layer weights.
$\hat{\b L}$ is a lower-triangular matrix with positive diagonal.
Therefore, it is a Cholesky factor and $\hat {\bm \Sigma}:=\hat{\b L}\hat{\b L}^\top$ obtains a valid covariance matrix.
The neural network outputs thus parameterize $p(\bm \theta \mid \b x)$.
By outputting a variance, the network directly models the signal-to-noise ratio of a data point.
This allows for filtering data points with too low signal-to-noise levels.

$\b W$ is obtained by training the network, which is done as follows.
We have a dataset $\mathcal{D} = \{(\b x_k, \bm \theta_k)\}_{k=1}^K$ of $K$ independent data-points, we obtain the optimal network weights using maximum likelihood.
\begin{align}
    \b W^*&:= \argmax_{\b W}  \log p(\mathcal{D} \mid \b W) \\
    &= \argmax_{\b W} \log \prod_{k=1}^K p(\bm \theta_k \mid \b x_k, \b W) \\
    &= \argmax_{\b W} \sum_{k=1}^K \log \mathcal{N}_{\b W}(\bm \theta_k \mid \hat{\bm \mu}_k, \hat{\bm \Sigma}_k)) \\
    &= \argmin_{\b W} \sum_{k=1}^K \mathcal{L}_k
\end{align} 
with loss function
\begin{equation}
    \mathcal{L}_k := \frac12 \log \det (\hat{\bm \Sigma}_k) + \frac12 (\bm \theta_k - \hat{\bm \mu}_k)^\top \hat{\bm \Sigma}^{-1}_k (\bm \theta_k - \hat{\bm \mu}_k) + c
\end{equation}
where $c$ is a constant that does not depend on $\b W$.
Using mini-batch gradient descent \citep{lecun2015deep}, we iteratively adapt $\b W$ to minimize this loss function. 
We use the Adam optimizer \citep{kingma2014adam} with default settings (e.g., learning rate) to do so.
Thereby, the neural network ``learns'' to predict the correct $\bm \theta$ from an input $\b x$.


\section{Results}
\label{sec:results}
We conduct several experiments to assess the efficacy of our methods.
The following are run on an Intel Xeon Gold 5118 with a maximum clock speed of 3.20 GHz and an NVIDIA Titan RTX GPU accelerator.

\subsection{Accuracy of Source Finding}
\label{sec:res_sf}
\begin{figure}[ht]
    \centering
    \includegraphics[width=0.8\linewidth]{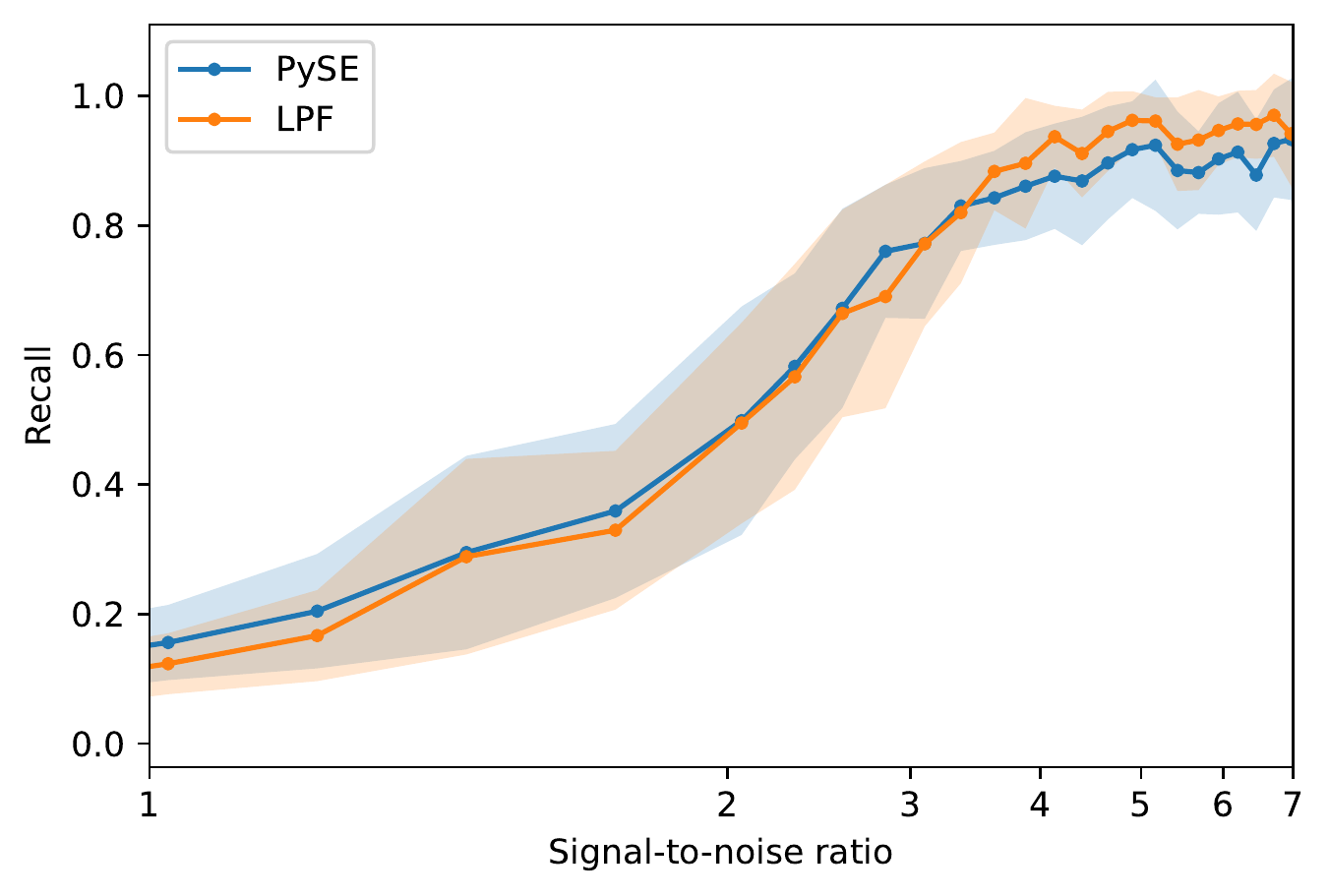} 
    \includegraphics[width=0.8\linewidth]{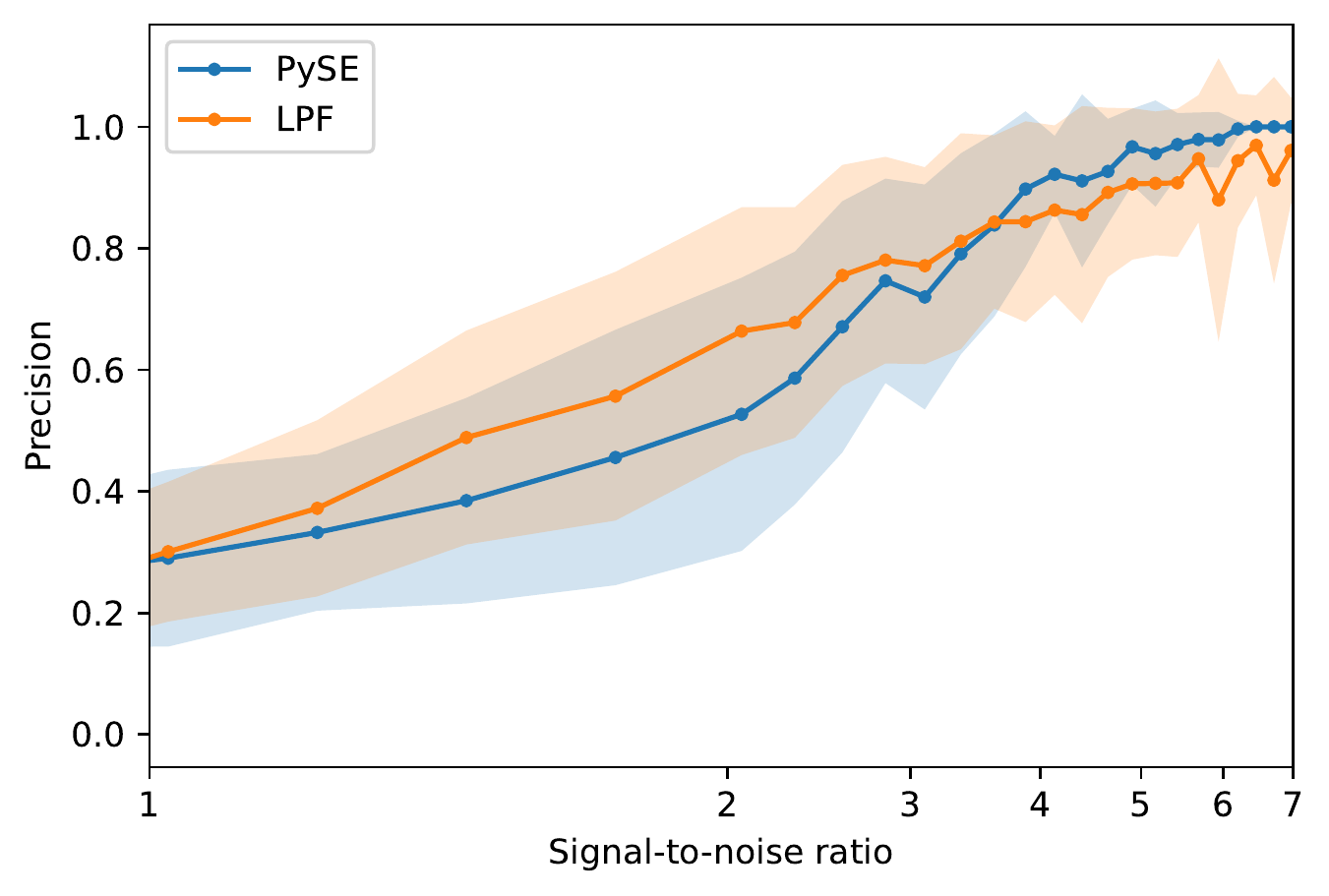}
    \includegraphics[width=0.8\linewidth]{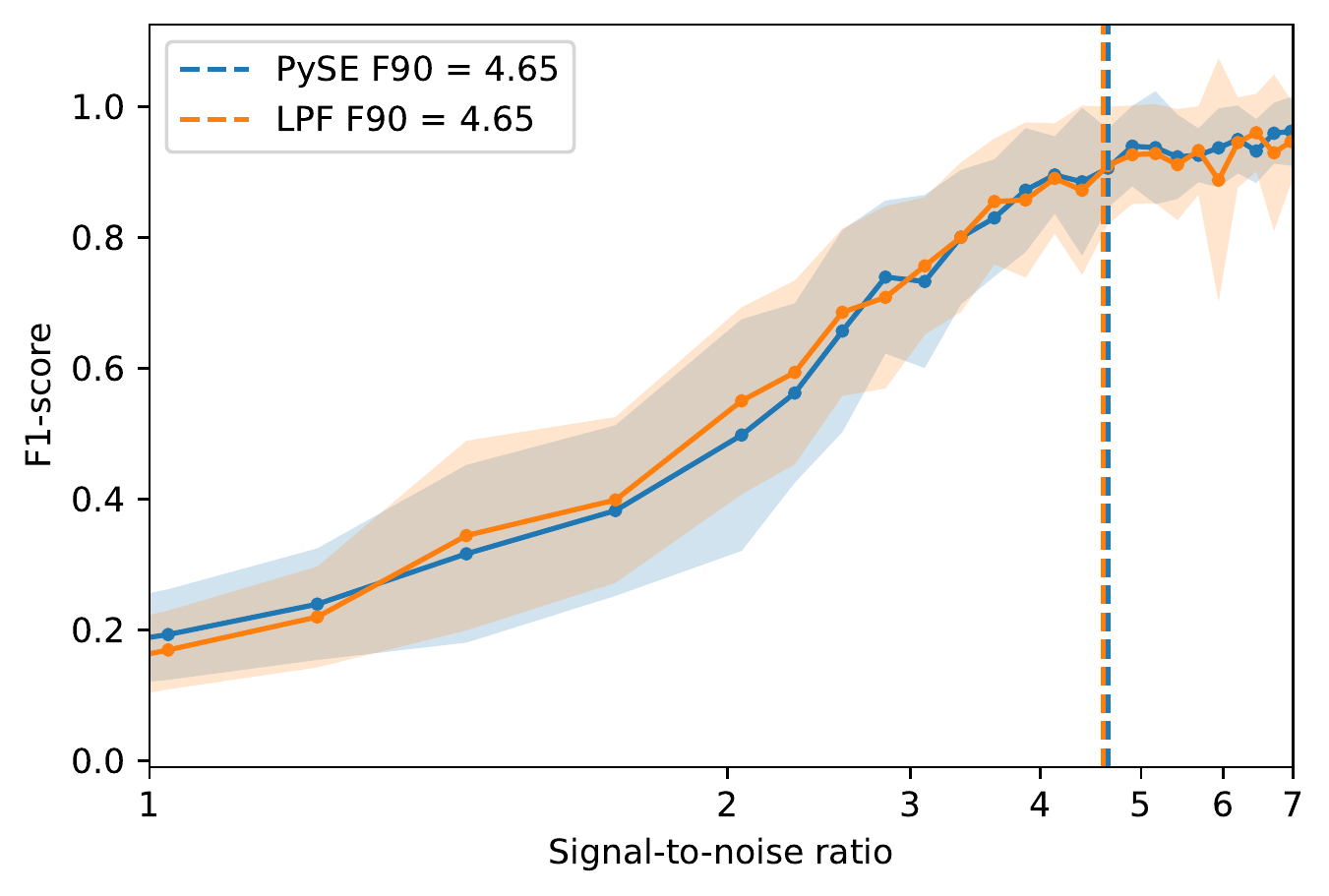}
    \caption{Recall, precision and F1 scores (higher is better) for our method (LPF) compared to PySE \citep{spreeuw2018pyse} as a function of signal-to-noise ratio. We report the 1-sigma bands of 32 runs. Additionally, we supply $F_{90}$ scores (lower is better) for both.}
    \label{fig:sf_results}
\end{figure}

To test the source-finder, we carry out experiments using single-bandpass all-sky images.
The images are simulated.
This has the advantage that we control entirely the noise characteristics and SNR of the sources.
First, we generate a model that simulates large-scale inhomogeneities (extended emission) of the final image. 
For example, our Galactic foreground noise and Cygnus A \& Cassiopeia A calibration remnants.
Then, we add to this map standard, independent Gaussian noise.
Following \cite{vafaei2019deepsource}, we sample flux values from an exponential distribution such that approximately 40\% have a signal-to-noise ratio less than unity.
We create point spread functions (PSF) by adding artificial side-lobes to a Gaussian main beam.
More detail and examples are shown in \cref{sec:supp_sky_simulation}.

We also take inspiration from \cite{vafaei2019deepsource} evaluating the source-finder.
The reported signal-to-noise ratios (SNRs) ($x$-axis) are computed by dividing the ground-truth peak flux value over the image noise.
We bin them and report the scores as a function of these bins.
The scores are precision (purity), recall (completeness), and F1.
\begin{equation}
    \text{P} := \frac{\text{TP}}{{\text{TP} + \text{FP}}}
\end{equation}

\begin{equation}
    \text{R} := \frac{\text{TP}}{\text{TP} + \text{FN}}
\end{equation}
\begin{equation}
    \text{F1} := \frac{2\text{PR}}{(\text{P} + \text{R})}
\end{equation}
where $\operatorname{TP}$ and $\operatorname{FP}$ are the number of true positives and false positives, respectively.
The F1 score computes a harmonic mean between precision (P) and recall (R). 
This is important as usually there is a trade-off between them.
We also denote what we call the $F_{90}$ which is defined as the minimal signal to noise ratio (SNR) such that the F1 score is at least $0.9$.
We run the methods at $\kappa=2$, which optimized the reported $F_{90}$.
For our method (coined ``LPF'' for Live Pulse Finder) we ran only a single sigma-clip iteration since this already recovers most of the true positives (see \cref{sec:supp_sigmaclip_iterations}).
In practice, we run more sigma-clip iterations as this maximizes recall.
In Figure \ref{fig:sf_results}, we compare our results with PySE \citep{spreeuw2018pyse} at various signal-to-noise ratios.
At similar recall values, our method is a bit more precise at low signal-to-noise ratios but slightly less precise with high signal-to-noise ratios.
In terms of F1 scores and the methods perform similarly with an $F_{90}$ of $4.65$.
Note that PySE takes around 2 seconds to complete the analysis for one image (single band) on our system (see \cref{sec:scaling}).

\subsection{Parameter Inference}
\label{sec:res_pi}
As specified in \cref{sec:parameter_inference}, we need a dataset of input-output pairs to train our neural network.
Since we have few true transient candidates, we are very short on such data.
Favorably, we know one of the most important properties of a true astrophysical transient: a bright, broadband, dispersed signal. 
The dispersion of the signal is manifested through the broadening of the pulse over a finite bandwidth.
It originates from the interaction of emitted photons with electrons along the path between an observer and the source.
The integrated electron column density, called the DM, is used as a proxy for the distance to the source.
Using known equations, we can easily simulate dispersed astrophysical transients.
The dispersion measure of non-astrophysical transients usually is close to 0 (i.e., no broadening of the pulse), making it possibly the most useful feature for separating signal from noise \citep{kuiack2020apparent}.
We build a dataset by injecting these simulated pulses into randomly sampled noise from the survey.
In the following, the reported parameter values can be set according to the interests of the practitioner.
They should be set such that the simulated dataset covers the population of interest sufficiently.
First, we generate a Gaussian pulse profile using a width $w$ sampled uniformly between $0$ and $16$, corresponding to a maximum FWHM of $37.67$ time-steps.
The profile is computed as 
\begin{equation}f(t) = \exp\left[-\left(\frac{t-t_0}{w}\right)^2\right]\end{equation} centered around $t_0$.
For each frequency $\nu$ of the survey, we compute the arrival time by applying a dispersion delay
\begin{equation}
t_\nu := t_0 + C_{\text{DM}} \text{DM}\left(\nu^{-2} - \nu_0^{-2} \right)
\end{equation}
with $C_{\text{DM}}$ the dispersion constant $ \frac{e^2 \text{pc}}{8 \pi^2 \epsilon_0 m_e c} \cdot 10^{-6}\approx 4148.806$  $\text{MHz}^2 \text{cm}^{-3} \text{ms}$ \citep{lorimer2012handbook}.
In this analysis, we sample the $\text{DM}$ uniformly between $0$ and $512$, covering a significant part of the known FRB population \citep{petroff2016frbcat}.
The final intensity is computed as
\begin{equation}
    I(t, \nu) = A\left(\frac{\nu}{\nu_0}\right)^\alpha f(t)
\end{equation}
where $A$ is the burst amplitude in standard deviations of the noise.
We experimented with $A$ uniformly between $0$ and $8$ and $\alpha$, the spectral index, sampled from $(-4, 4)$ uniformly.
We compute the intensity for all time-steps and frequencies and save the pulse at the corresponding band and time indices.
Using this procedure, we create $32,768$ samples (input time-frequency plots and output parameters pairs) to train the network.

We train the convolutional neural network (CNN) on the generated dataset.
The network was trained using early stopping \citep{lecun2015deep}.
That is, training was not canceled until the likelihood of the held-out validation data stopped increasing.  
\begin{table}
\centering
\small
\setlength{\tabcolsep}{3pt} 
\renewcommand{\arraystretch}{1.5} 
\begin{tabular}{@{}cc|cccc@{}}
\cmidrule(l){2-6}
                                & Fluence   & (0.0, 1.0]    & (1.0, 2.0]      & (2.0, 4.0]     & (4.0, 8.0] \\ \cmidrule(l){2-6} 
                                & MAE       & 119.3 & 28.19 & 13.25  & 7.982  \\
                                & MAE/DM       & 2.868 & 0.750 & 0.170  & 0.053    \\
                                & RMSE      & 165.4 & 48.06 & 23.14  & 16.07      \\
                                & RMSE/DM      & 10.77 & 7.612 & 1.162  & 0.182 \\
\multirow{3}{*}{\rotatebox[origin=c]{90}{\scriptsize{True value within}}}
                                & $1\sigma$ & 0.590 & 0.650 & 0.635 & 0.641  \\
                                & $2\sigma$ & 0.917 & 0.955 & 0.923 & 0.952  \\
                                & $3\sigma$ & 0.983 & 0.995 & 0.992 & 0.999 \\ \cmidrule(l){2-6} 
\end{tabular}
\caption{Quantitative results of the neural network prediction of the DM parameter. The column values are fluence bins. We report the mean absolute error (MAE), root-mean-square error (RMSE; lower is better) absolutely and relative to the dispersion measure. Additionally, we give the probabilities that the true value lies within 1, 2 or 3 times the predicted standard deviation.}
\label{tab:results}
\end{table}
Quantitative results on the dispersion measure inference (arguably, the most important feature for discriminating spurious from real bursts) are shown in \cref{tab:results}.
The reported MAE (mean absolute error) and RMSE (root-mean-squared error) are satisfactory considering the range of values that the dispersion measure can take, which is confirmed by observing the relative MAE/DM and RMSE/DM.
The uncertainty is relatively well-calibrated. 
That is, it is close to what one expects for a Gaussian distribution.
Notably however, the network is overconfident for low-fluence bursts.


\subsection{Testing the Pipeline}
\begin{figure}
    \centering
    \includegraphics[width=0.8\linewidth]{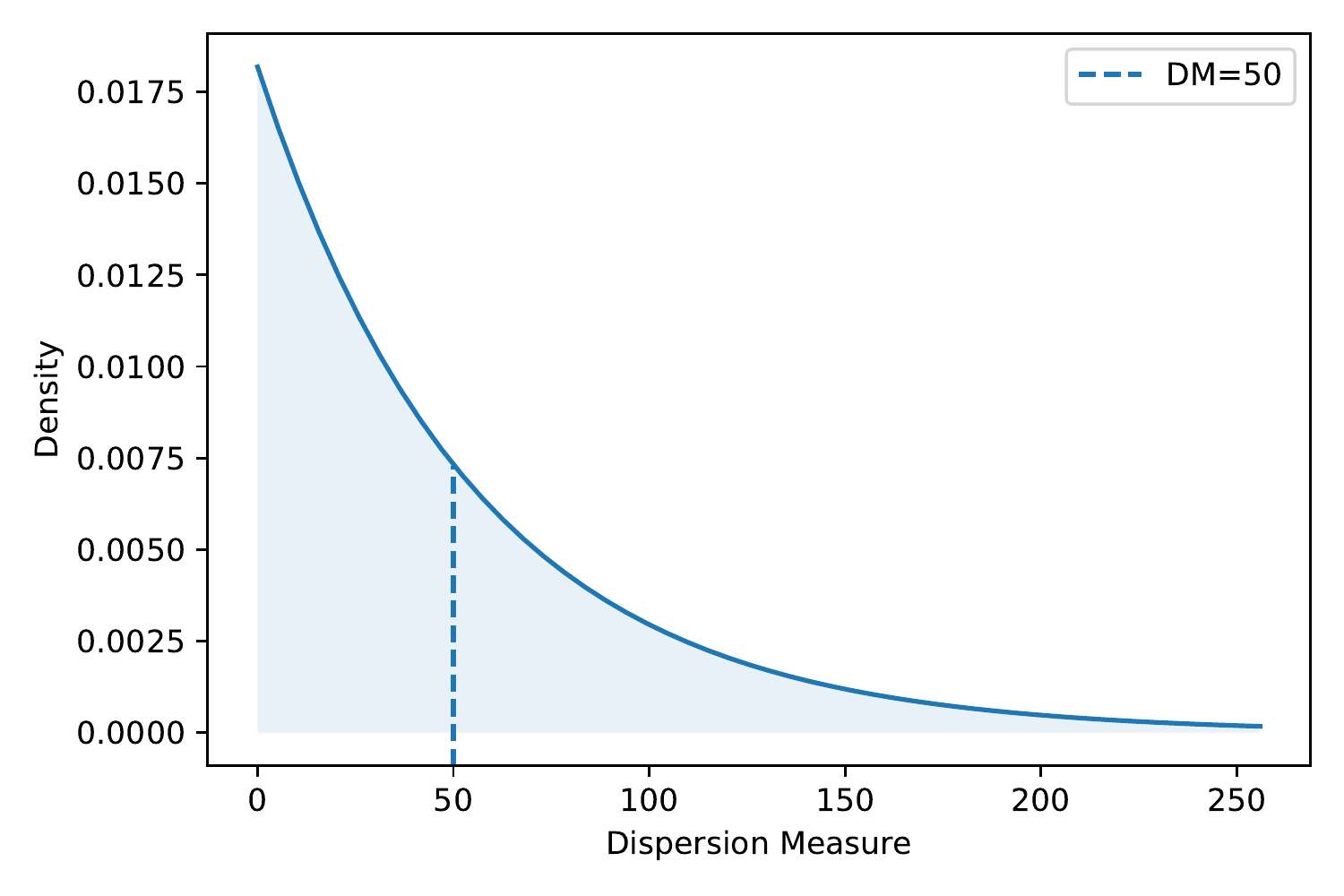}
    \caption{Population density from which the dispersion measure was sampled for the simulated transients in our all-sky pipeline testing experiment. Roughly 40\% of the samples have a dispersion measure DM $>50$.}
    \label{fig:dm_density}
\end{figure}
We now test the efficacy of the entire pipeline.
We simulate entire sky images consisting of both stable sources and (dispersed) transients.
We sample the signal-to-noise ratios for both the transients and the stable sources using the same exponential distribution as in \cref{sec:res_sf} (i.e., as suggested by \cite{vafaei2019deepsource}).
The dispersion measure was sampled from an exponential distribution with rate parameter $\frac{1}{55}$, such that roughly 40\% of the transients have a dispersion measure $\mathrm{DM}>50$ (\cref{fig:dm_density}).
This is somewhat arbitrary and partly set such that there are sufficient dispersed signals, but can be motivated as follows:
First, the majority of detections are expected to be lowly dispersed (atmospheric) noise.
Additionally, the integrated flux within a passband is reduced due to temporal dispersion.
We, therefore, do not expect many highly dispersed pulses to be detectable at all.
Finally, at the long wavelengths of AARTFAAC \citep{prasad2016aartfaac}, extremely dispersed events are so smeared out that the slope would not be inferrable from the dynamic spectrum.
For inference, the same neural network is used as in \cref{sec:res_pi}.
Note that this network was trained on uniform parameter distributions, meaning that there is a discrepancy between the training data and the data that we test the model on.
This replicates the distribution shift we would also expect in real data.
Thus, we first inspect how the recovered population of transient candidates matches the ground truth one.
Thereafter, we assess how many of the transients are recovered (as a function of fluence and dispersion measure).

\begin{figure}
    \centering
    \includegraphics[width=0.8\linewidth]{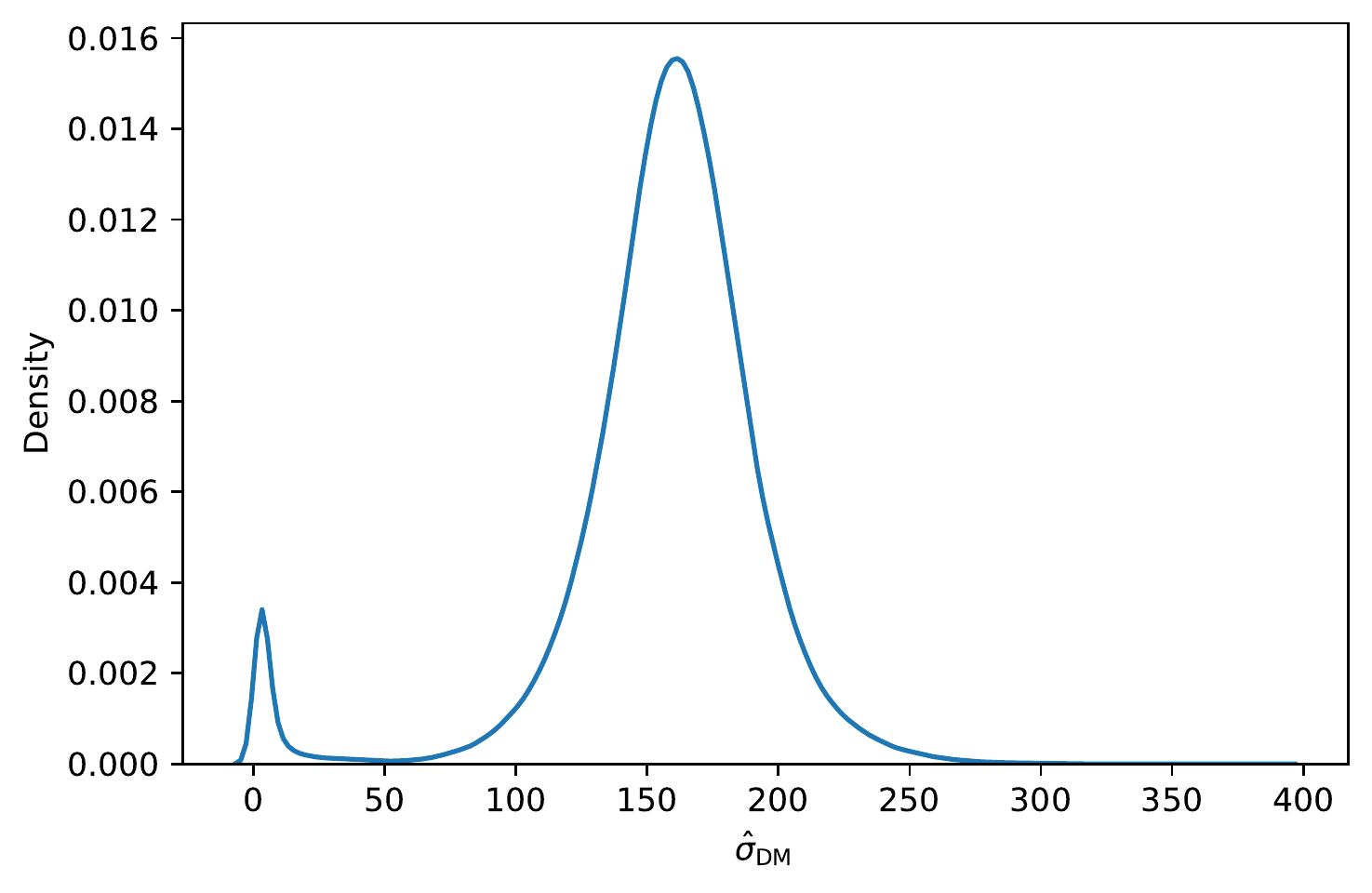}
    \caption{Density plot of the by the model output standard deviation of the dispersion measure $\hat{\sigma}_{\mathrm{DM}}$. We observe a clear bimodal distribution. The high uncertainty mode corresponds to false positives. }
    \label{fig:model_dm_std}
\end{figure}
\begin{figure}
    \centering
    \includegraphics[width=0.8\linewidth]{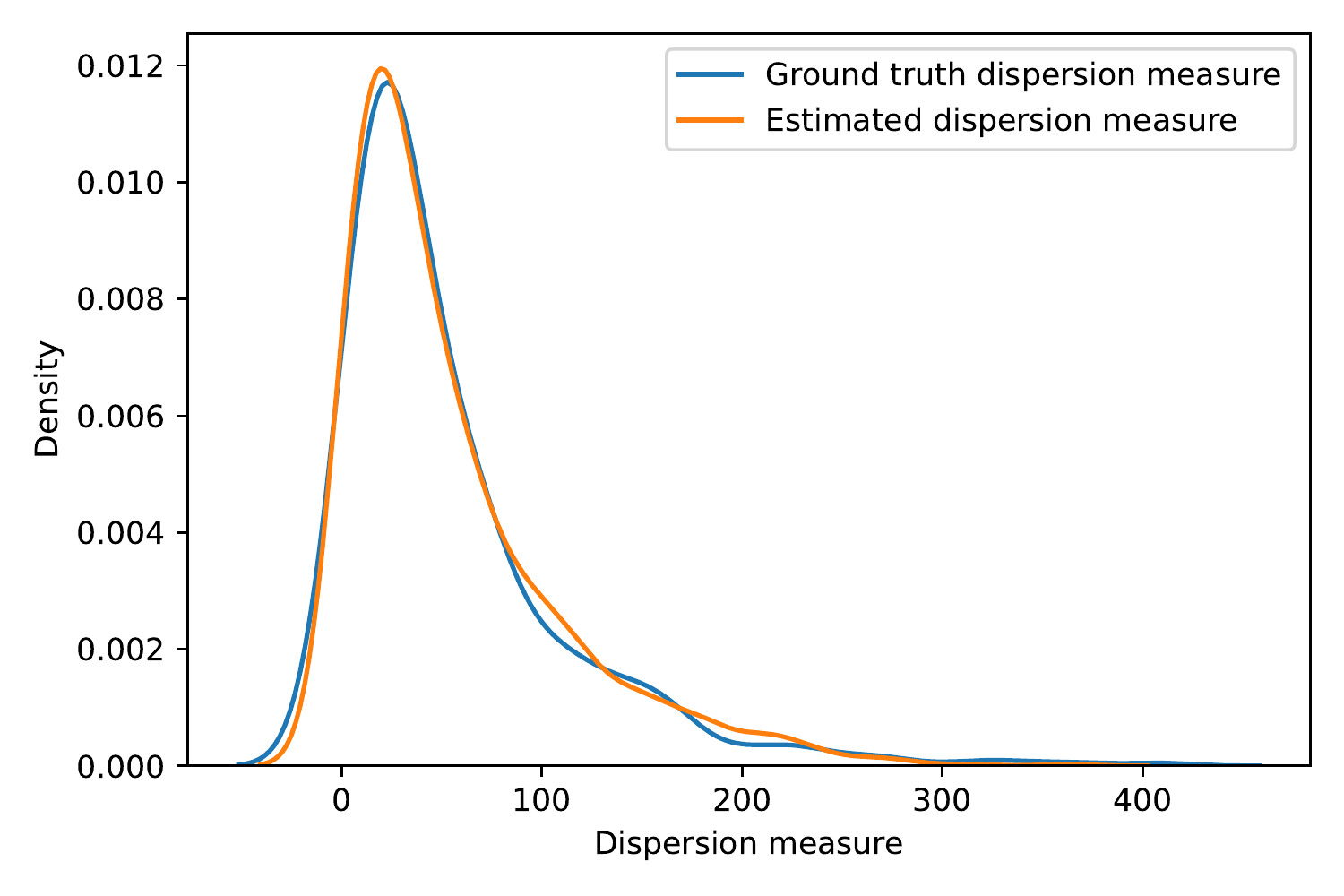}
    \caption{Sample (simulated) density and recovered density of transient dispersion measures. Note that the model was trained on a different density (uniform), but is not affected much by the distribution shift.}
    \label{fig:recovered_densities}
\end{figure}
\begin{figure}
    \centering
    \includegraphics[width=0.8\linewidth]{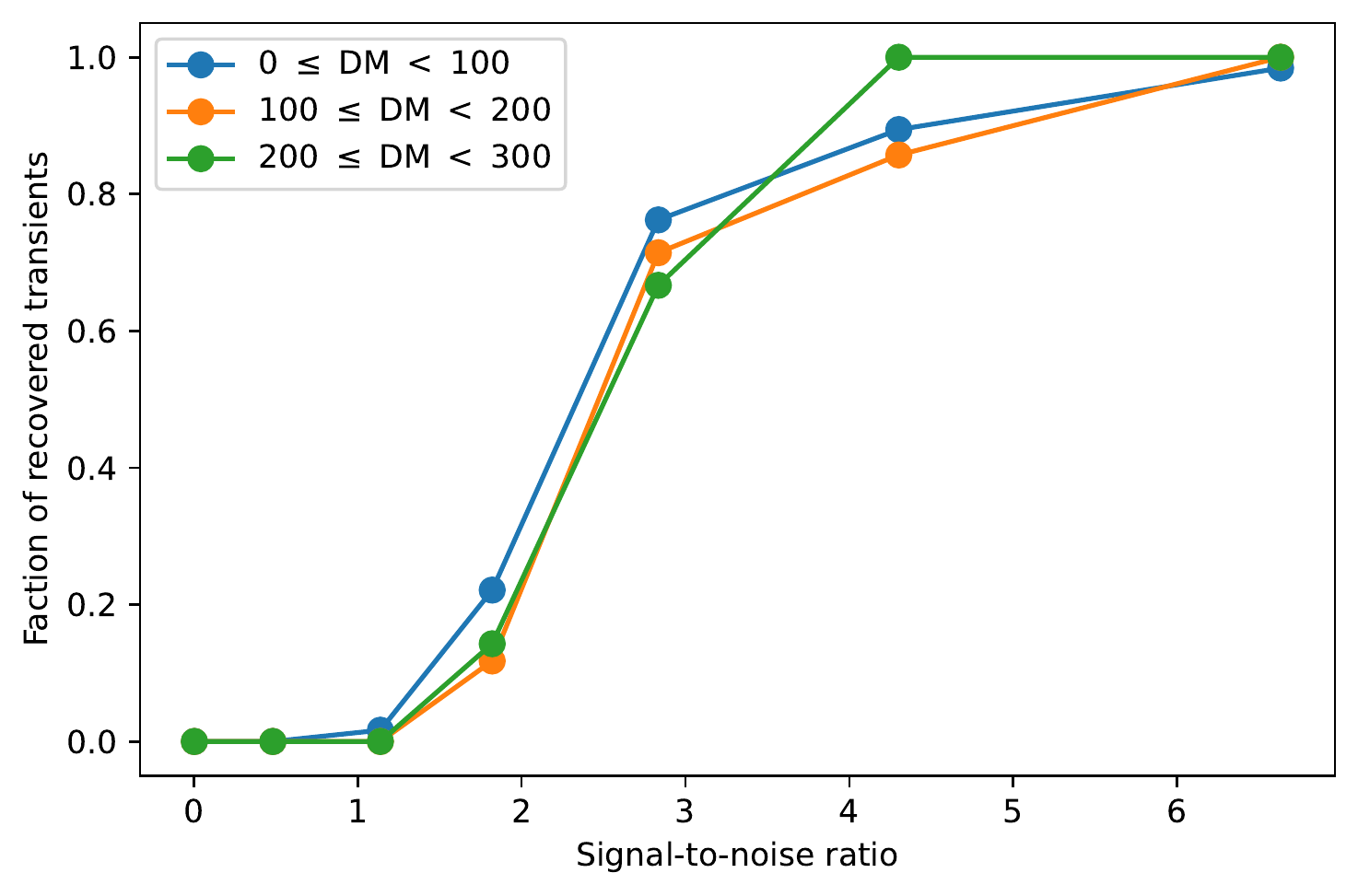}
    \caption{Fraction of recovered transients as a function of signal-to-noise ratio (higher is better) averaged in dispersion measure (DM) bins.}
    \label{fig:recall_to_snr}
\end{figure}

First observe \cref{fig:model_dm_std}, where we plot the density of modelled standard deviation for the dispersion measures ($\hat{\sigma}_{\mathrm{DM}}$, found on the diagonal of $\hat{\b \Sigma}_k$) of the sources. 
We clearly see a bimodal distribution.
The high uncertainty modes correspond to detections that were simply noise or stable sources.
After removing the high uncertainty mode $\hat{\sigma}_{\mathrm{DM}} > 50$ (regarding them false positives), we plot the resulting dispersion measure density in \cref{fig:recovered_densities}.
Even though the model is trained on uniform distributions, the recovered predictions closely follow the expected density.
Note that we do not expect the network to infer parameters that are extremely far from its training distribution (e.g., $\mathrm{DM} \gg 512$) correctly.
However, the recovered distributions are not very much biased to the uniform distributions that the model was trained on, which is reassuring.
In \cref{fig:recall_to_snr} we show the percentage of recovered transients as a function of signal-to-noise ratio for three different dispersion measure bins.
As expected, the fraction of recovered transients approaches unity with increasing signal-to-noise ratios.
\begin{figure*}[ht]
    \centering
    \includegraphics[width=0.26\linewidth]{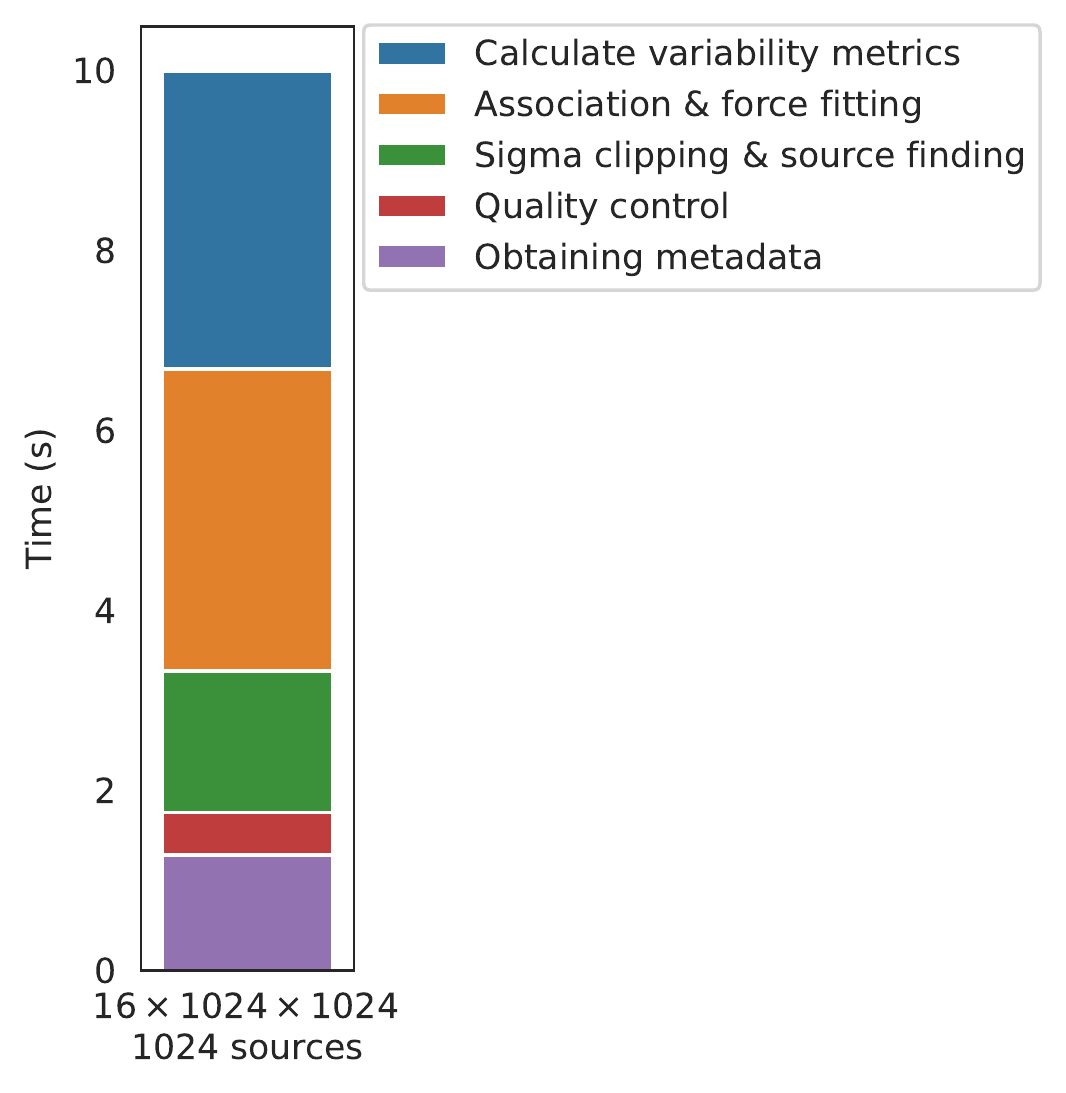}
    \includegraphics[width=0.4\linewidth]{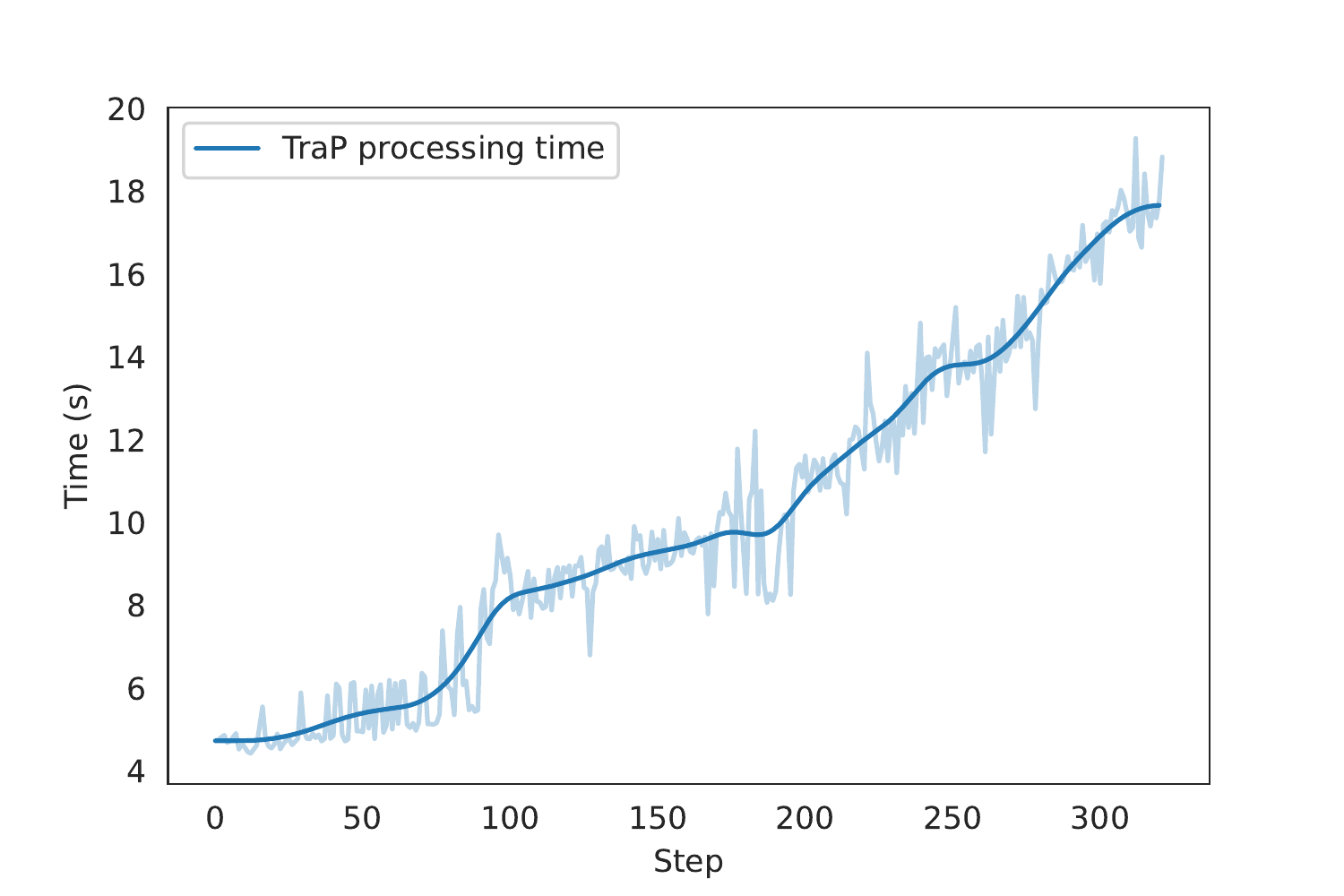} \\
    \includegraphics[width=0.55\linewidth]{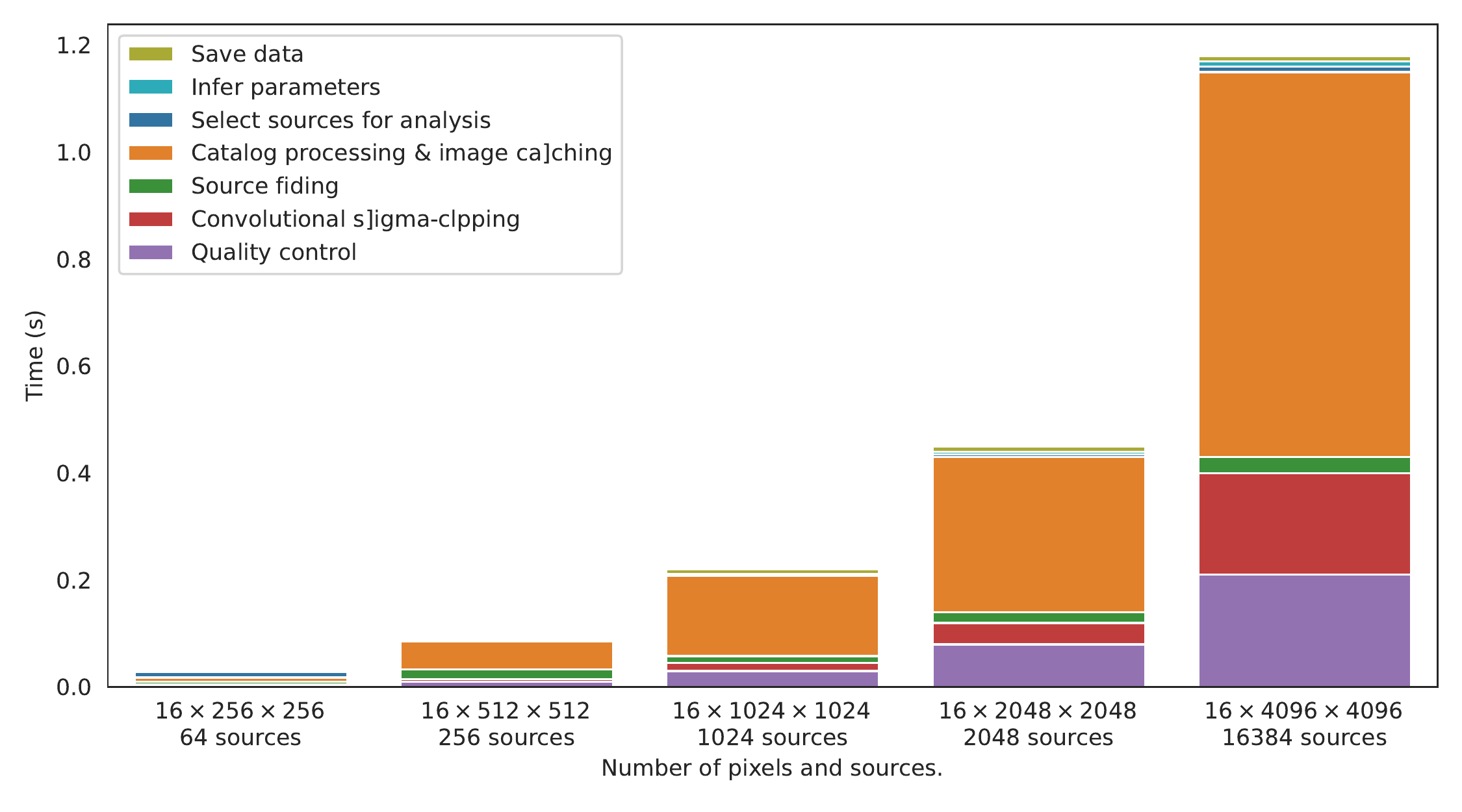} 
    \caption{Left: Timing of the \cite{swinbank2015lofar} pipeline for $16$ $1024 \times 1024$ images. Reported numbers are averages of the timings as reported in the right figure. Right: due to the extensive association procedure and variability metrics calculation, the time taken per image scales superlinearly and thus is prohibitive for lengthy observations. Bottom: scaling of the entire pipeline and the individual steps for different image sizes and source numbers. Note that in contrast to \cite{swinbank2015lofar}, our method does not scale with the number of time-steps.}
    \label{fig:lpf_scaling}
\end{figure*}
\subsection{Application to Real Data}
We applied the pipeline to a real AAARTFAAC-6 survey dataset.
We used $\kappa=4.5$ to retrieve an acceptable number of candidate sources from the source detection pipeline.
``Acceptable'' here expresses a trade-off between precision and recall. 
By lowering $\kappa$, many more spurious candidates will be retrieved. 
The neural network will filter these, but the disadvantages are two-fold: more processing has to be done and it might start force-fitting at locations where you want to remain sensitive to actual candidates.
After pushing the associated time-frequency data through the network, we obtain a vector of parameters $\{\hat{\bm \theta}_l \}_{l=1}^L$ for $L$ recovered transient candidates.
These can be filtered based on the scientist's needs.
To obtain examples of interesting bursts presented in this study, we used the following.
We threshold the inferred dispersion measure $\hat{\mu}_{DM}$ at 50.
Next, we sorted the candidates according to their inferred dispersion measure standard deviation $\hat{\sigma}_{\mathrm{DM}}$ and inspected them top-down.
By applying the entire pipeline to a real-time live survey we retrieved interesting candidates for follow-up analysis fully automatically.
Examples are shown in \cref{fig:examples}.
The above is exemplary but gives an idea of how a practitioner can filter the data based on inferred physical parameters.
The proposed transient \cite{kuiack2020aartfaac} was recovered with an estimated dispersion measure of $74.49 \pm 8.26$ (\cite{kuiack2020aartfaac} concluded a dispersion measure of $74 \pm 5$).


\subsection{Scaling results}

\label{sec:scaling}
Finally, we report performance results as scalability was one of the research goals of this work.
In \cref{fig:lpf_scaling} we see that we can process images in a fraction of the time that \cite{swinbank2015lofar} takes.
Importantly, we can process $16\times 1024 \times 1024$ cubes in real time, which is a science goal of this work.
Also comparably, \cite{pintaldi2021scalable} report 924 images processed in 13 hours.
We see that the processing time as a function of the number of pixels scales sub-linearly.
There is little overhead added by our convolutional source localization: the bulk of compute is required for catalog processing and image caching.
Specifically, backward filling (see \autoref{sec:assoc}) takes 37\% of these steps, and 
image caching takes 60\%, increasing with image size.
Smart I/O can solve this but is left for future work.
Finally, inferring directly parameters using a neural network adds little overhead.

\section{Conclusion}
\label{sec:conclusion}
We presented new methods that allow for real-time analysis of all-sky radio image cubes on time scales where dispersion starts to play a role.
In it, parallelized methods based on convolutions and filters that are accelerated on a GPU process the image stream.
Afterward, a neural network is employed to infer physical parameters, among which the dispersion measure of the detected sources.
Based on these, false positives are easily filtered.
The methods were tested individually and as a whole on simulated data, as well as on real data from AARTFAAC, a LOFAR based transients facility.
The results can competently recover simulated dispersed transients from the simulated data stream.
In real AARTFAAC data dispersed signals were found, on which follow-up analysis can be performed.
Scaling results showed that the entire pipeline can analyze image cubes upwards of $16$ $1024\times 1024$ images in under $1$ second per iteration.
Thus, applying the method in real time can filter uninteresting data, providing a solution to the big-data problem that modern astronomy is dealing with.
Concluding, the current work proposed effective and efficient methods to search for intermediate-length dispersed transients in radio image cubes, filling a methodological gap that is also relevant for MWA, LOFAR, LWA1, OVRO-LWA and the future SKA.

\section{Future Work}
Based on the blind detections we performed so far, it is clear that just searching for broad-band dispersed signals is not enough. 
Scintillation seems to produce a significant fraction of the dispersed signals we are looking for, and we should find new ways to sift the distribution of these candidates from truly interesting events.
Some of the other parameters (e.g. the width of the pulse) seem promising.
Future research could further investigate.

\section{Acknowledgements}
This research made use of AstroPy \citep{price2018astropy}, Pandas \citep{mckinney2011pandas}, NumPy \citep{walt2011numpy}, SciPy \citep{virtanen2020scipy}, Matplotlib \citep{hunter2007matplotlib} and PyTorch \citep{paszke2017automatic}.
Accordingly, we would like to thank the scientific software
development community, without whom this work would not
be possible.
This work was fully funded by the University of Amsterdam.

\bibliographystyle{model2-names}
\bibliography{bib}







\appendix
\section{Quality Control}
\begin{figure}
    \centering
    \includegraphics[width=0.2 \paperwidth]{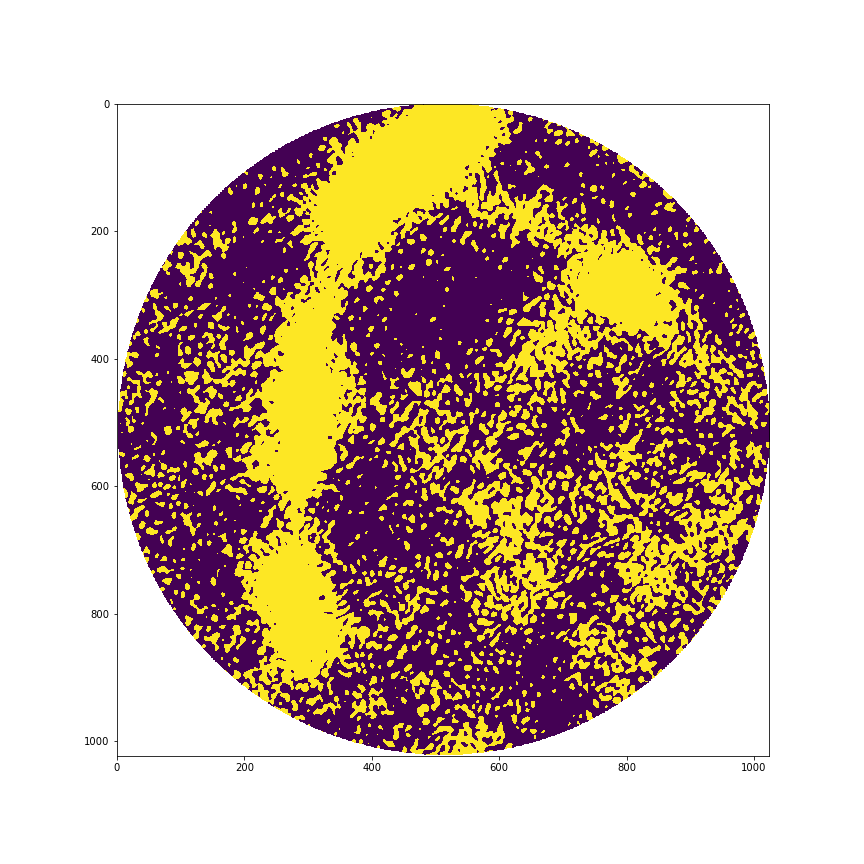}
    \caption{Example of resulting image of correlation and imaging pipeline error.}
    \label{fig:cor_example}
\end{figure}
\begin{figure}
    \centering
    \includegraphics[width=0.2 \paperwidth]{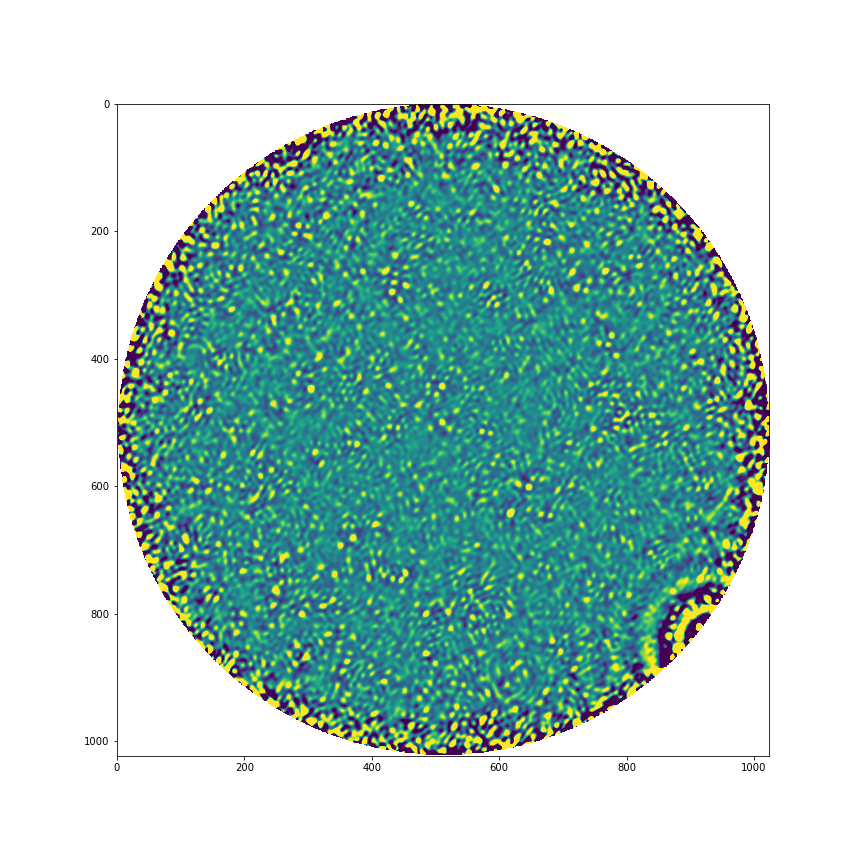}
    \includegraphics[width=0.2 \paperwidth]{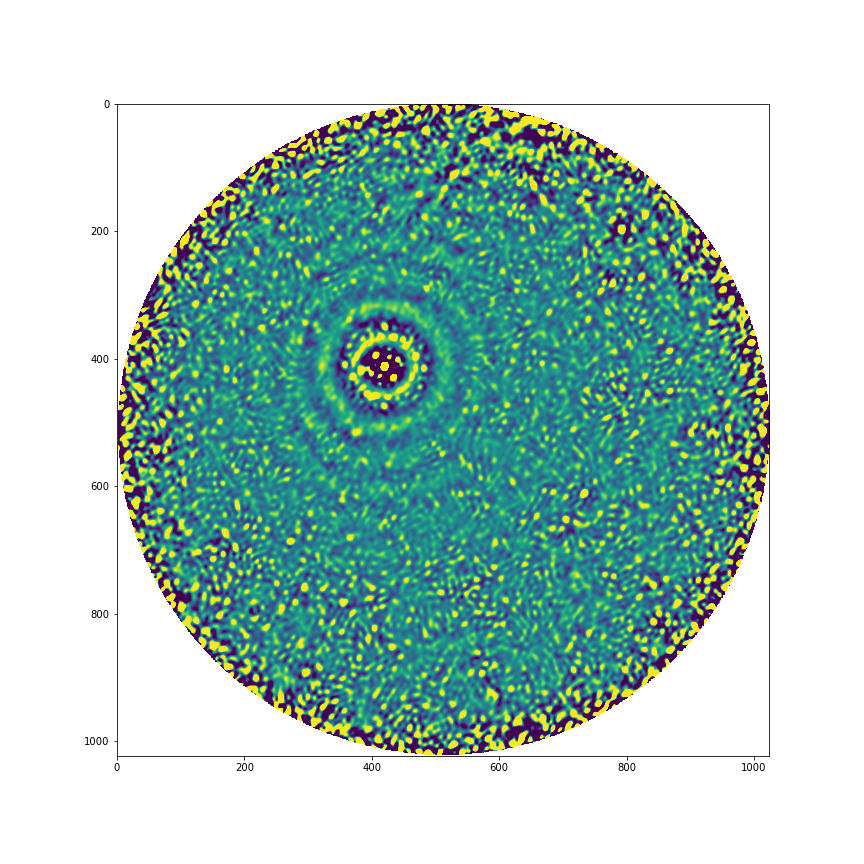}
    \includegraphics[width=0.2 \paperwidth]{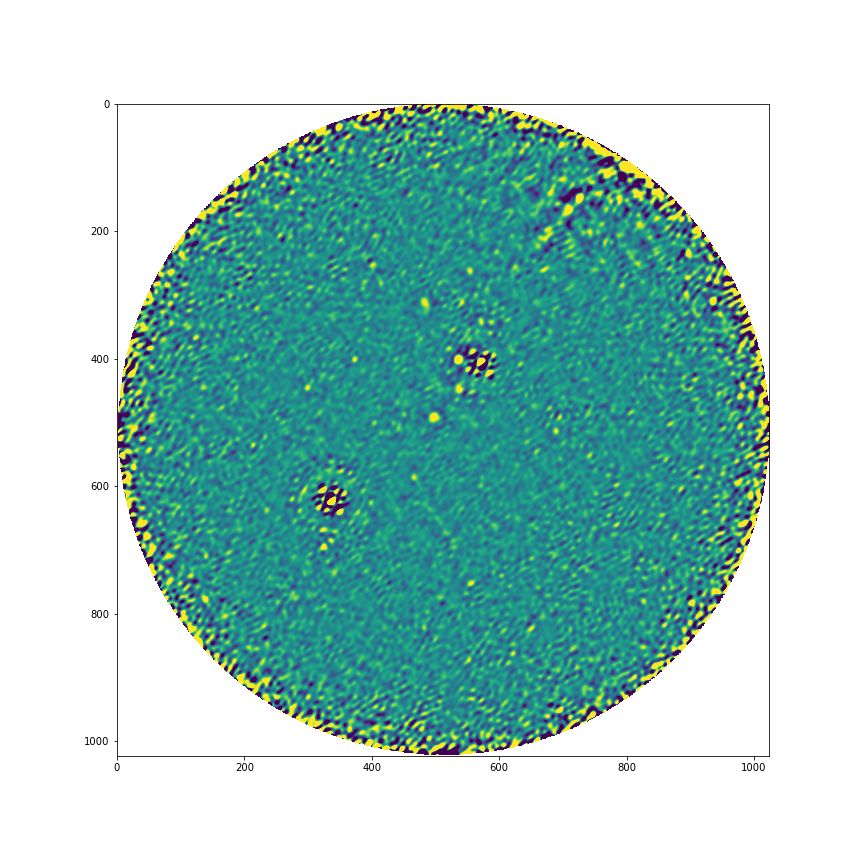}
    \includegraphics[width=0.2 \paperwidth]{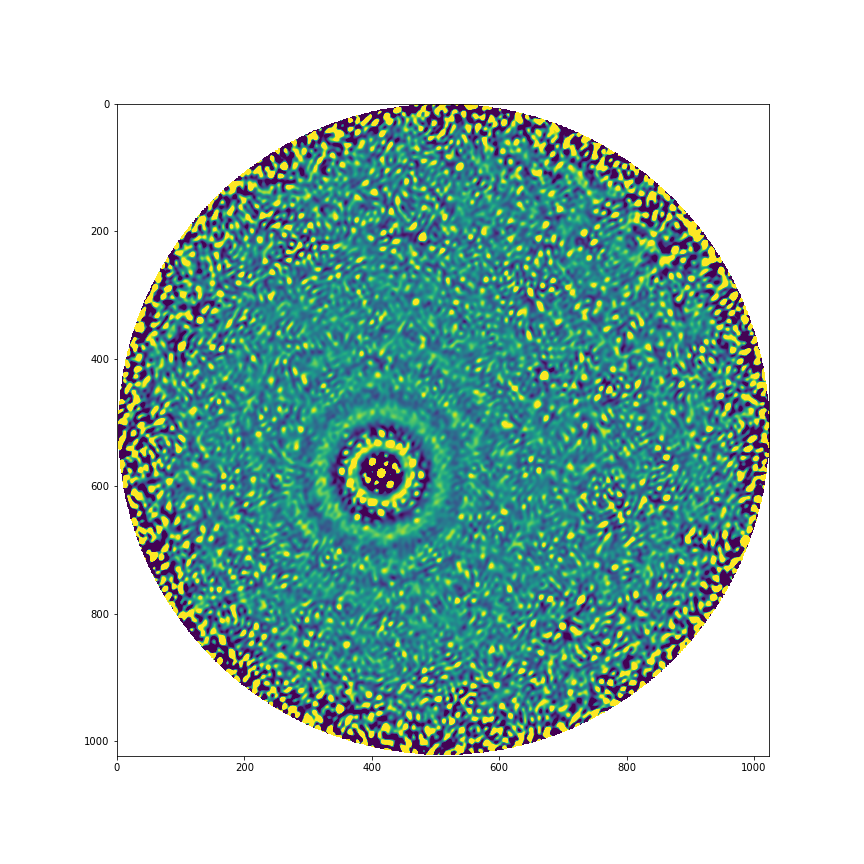}
    \caption{RFI corruptions detected using the proposed method. Note that these occurred only 1 sub-band adjacently to the clean counterpart presented in \cref{fig:a6_clean}.}
    \label{fig:examples_rfi}
\end{figure}
\begin{figure}
\centering
\includegraphics[width=0.45\linewidth]{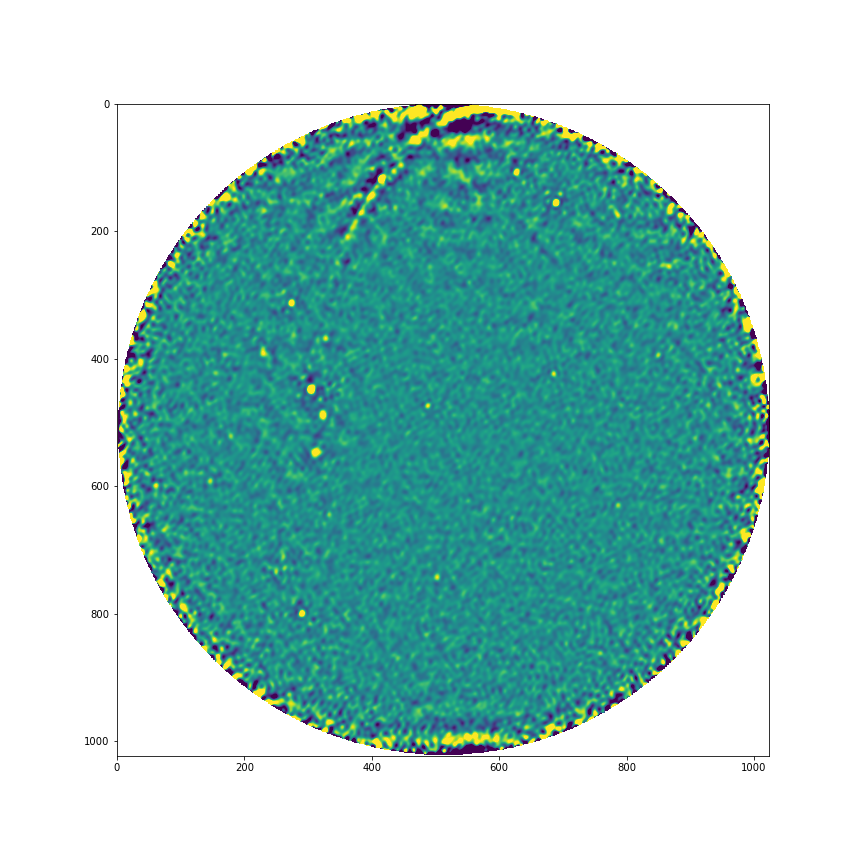} \includegraphics[width=0.45\linewidth]{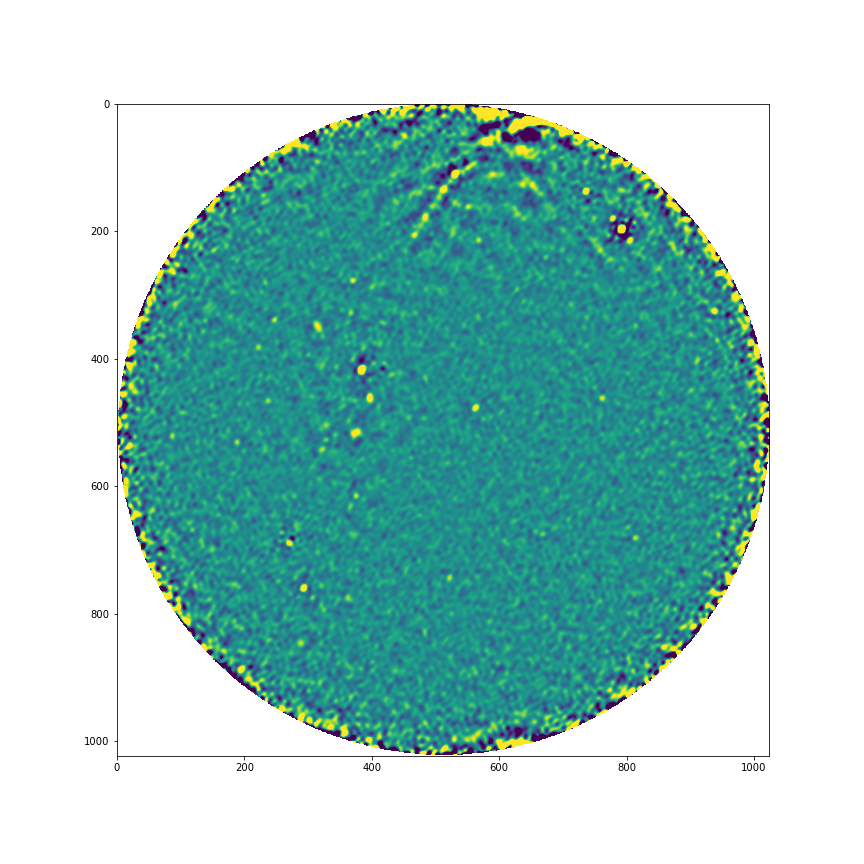}
\includegraphics[width=0.45\linewidth]{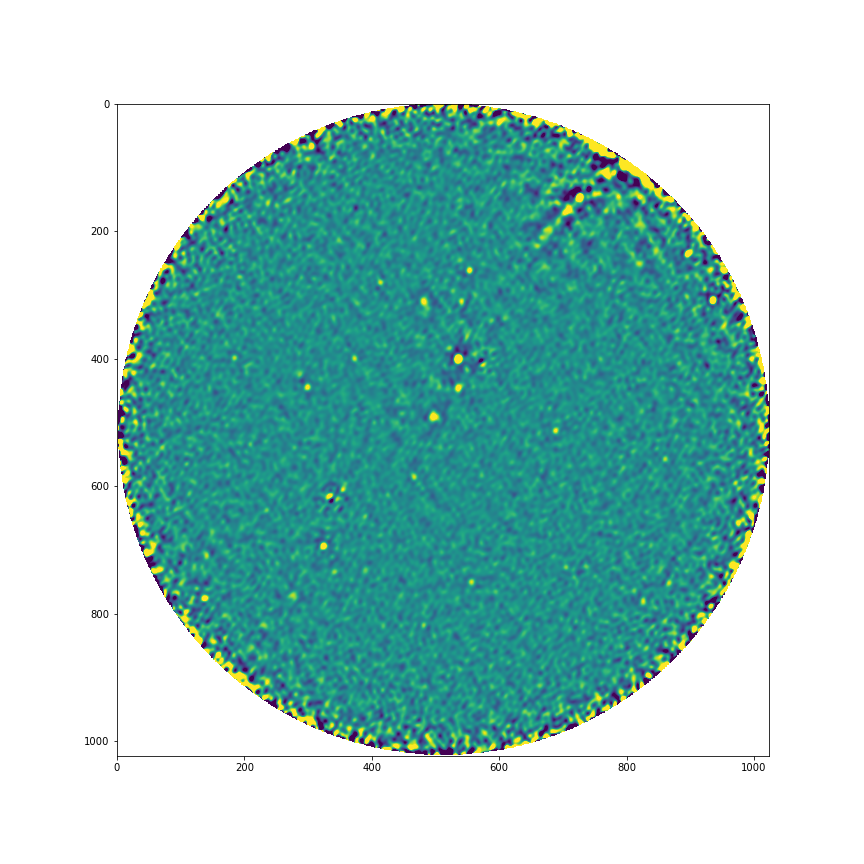}
\includegraphics[width=0.45\linewidth]{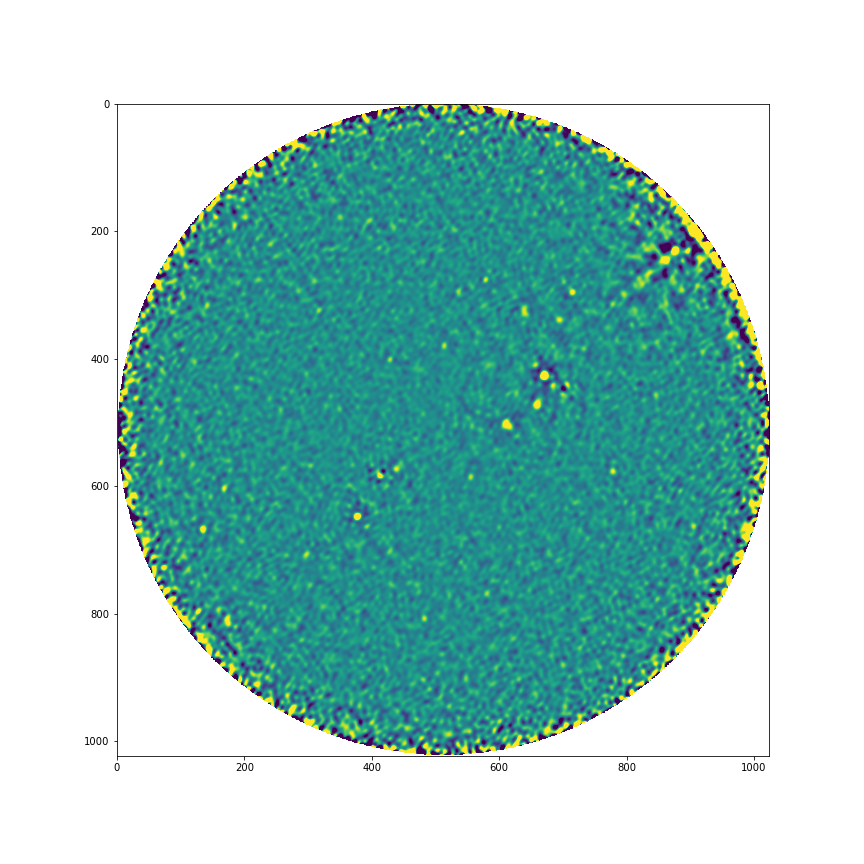}
\caption{Clean AARTFAAC6 images.}
\label{fig:a6_clean}
\end{figure}

\label{sec:supp_quality_control}
Before we let an image cube enter our pipeline, we make sure it is usable.
Not doing so leaves the downstream tasks vulnerable to e.g. numeric overflow or simply an intolerable number of false positives.
Moreover, attending to these false positives can cause true positives to be missed.
In the AARTFAAC data stream there are roughly two noise modes.
The first mode contains corruptions due to faults in the telescope or imaging pipeline.
These are subband-specific but are spatially global and last for extended periods of time.
An example is shown in \cref{fig:cor_example}.
The second mode consists of radio-frequency interferences that are usually local (spatial and frequential) and short-lived.
Examples are shown in \cref{fig:examples_rfi}.
To illustrate their frequential locality, the adjacent passband images are shown in \cref{fig:a6_clean}.
To detect outliers, one has to produce a center and scale of the data to compare to.
The telescope or imaging pipeline corruptions can last for extended periods and thus using temporal (moving) averages is futile, as the noisy data will be incorporated into the statistics.
Since we analyze data at multiple passbands, we can instead use the prior knowledge that these corruptions are frequentially local and thus use the other passbands to form center and scale estimates.
The corruptions usually are harsh, therefore we use robust statistics.
Consider an image cube at time $t$ (we omit the time index) $\b X \in \mathbb{R}^{B\times D\times D}$.
We take the average value of the band images in a vector $\b b := \frac1{D^2}\sum_{i=1}^D \sum_{j=1}^D \b X_{:, i, j}$.
We compute standardized robust scores
\begin{align}
    \b z := \frac{\b b - \mathrm{median}(\b b )}{\mathrm{MAD}(\b b)}
\end{align}
where MAD is the Mean Absolute Deviation \citep[e.g.][]{howell2005median}. 
A component $z_b$ of $\b z$ that exceeds a threshold (of e.g. 5) results in the corresponding image $X_b$ being discarded.
Discarding is done by imputing the entire image with zeros.

The radio-frequency interference spikes illustrated in \cref{fig:examples_rfi} are usually short-lived but spatially and frequentially more local.
Therefore, outliers cannot be detected in the averages of entire images.
We, therefore, keep running estimates $\bar{\b X}$ and $\bar{\b S}$ of the mean and standard deviation of the image stream using e.g. \cite{welford1962note, finch2009incremental}.
We compute pixel standardized scores
\begin{equation}
    \zeta_{bij} := \frac{x_{bij} - \bar{x}_{bij}}{s_{bij}}
\end{equation}
and if $\zeta_{bij}$ exceeds a threshold we discard the image $X_b$ in band $b$ by imputing it with zeros.
Note that the transients we are searching for could also exceed such a threshold.
However, observing \cref{fig:examples_rfi} we note that the radio-frequency interference spikes are much harsher than an astronomical transient.
Thus, we can set a $\zeta_{bij}$ such that only the extreme outliers are caught.
\section{Statistical Analysis of Source-finding}
\begin{figure*}[ht]
    \centering
    \includegraphics[width=0.3\linewidth]{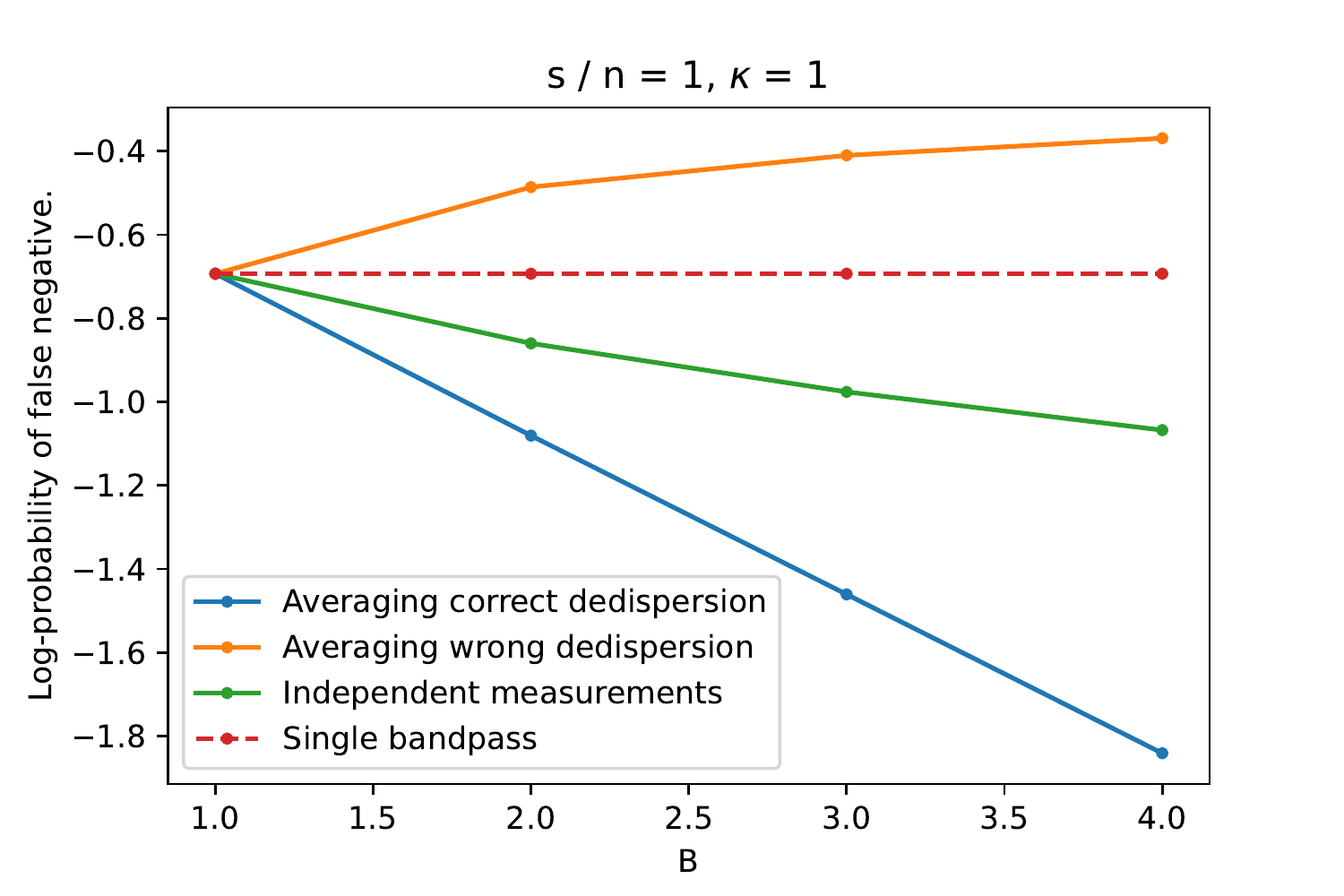}
    \includegraphics[width=0.3\linewidth]{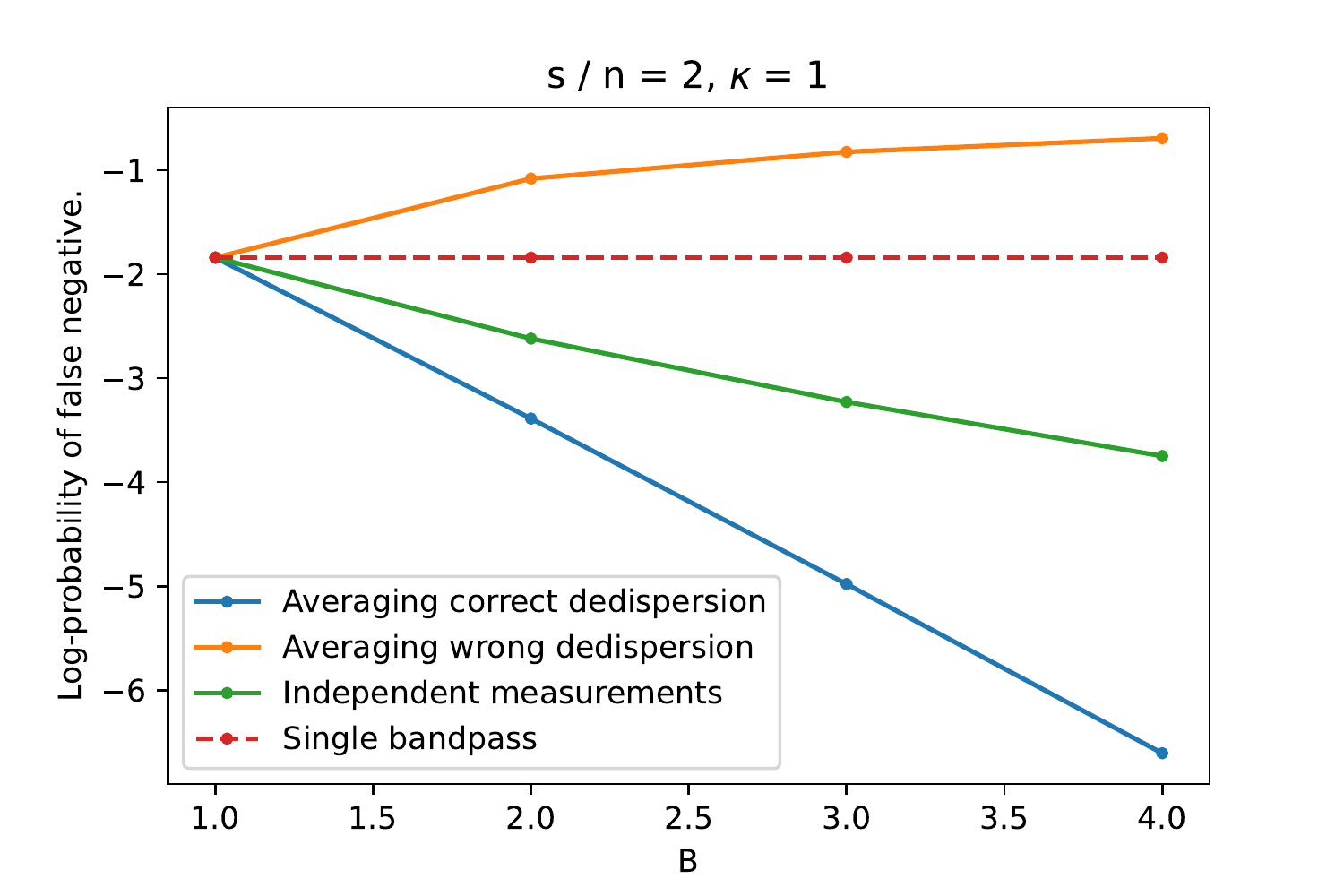}
    \includegraphics[width=0.3\linewidth]{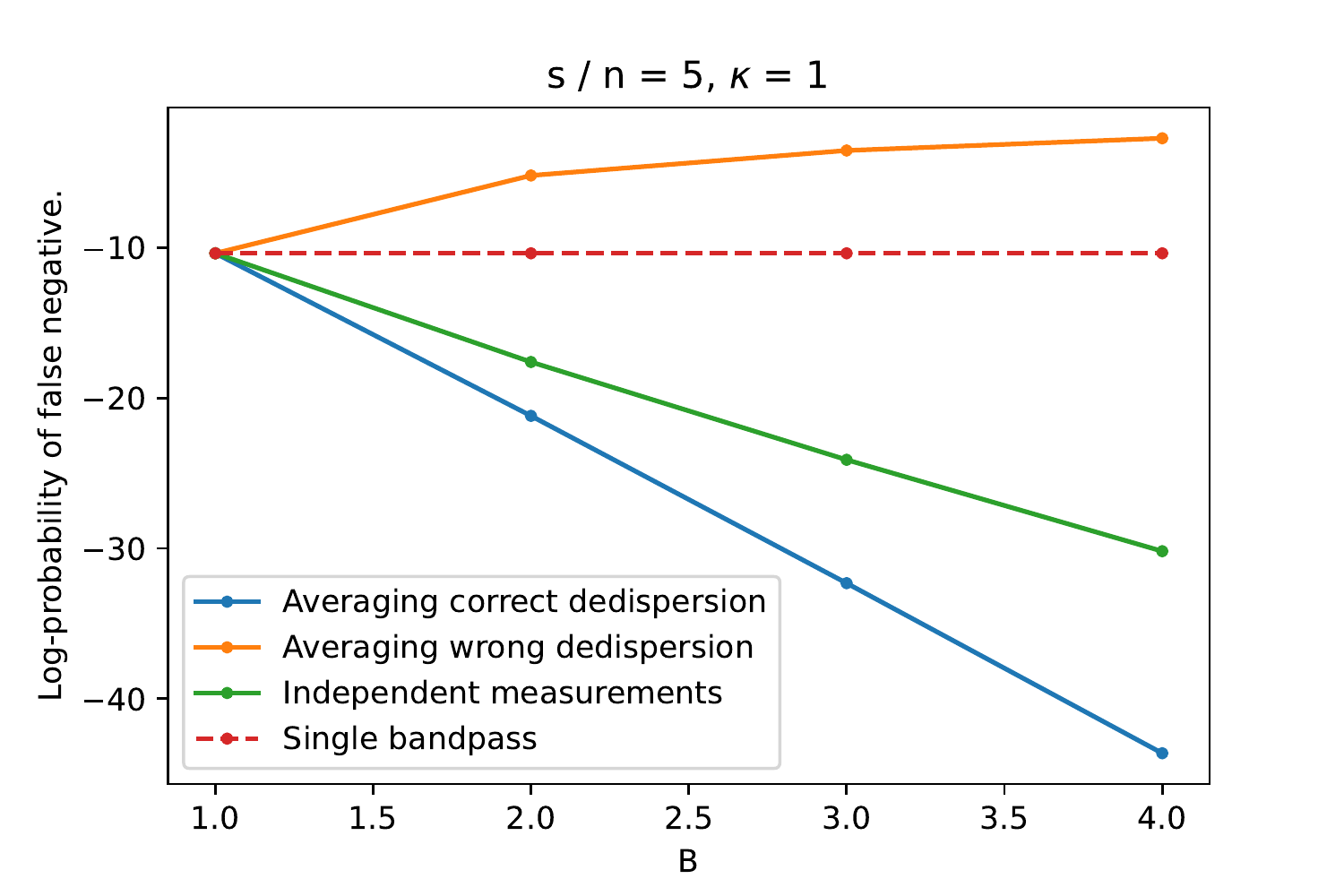} \\
    \includegraphics[width=0.3\linewidth]{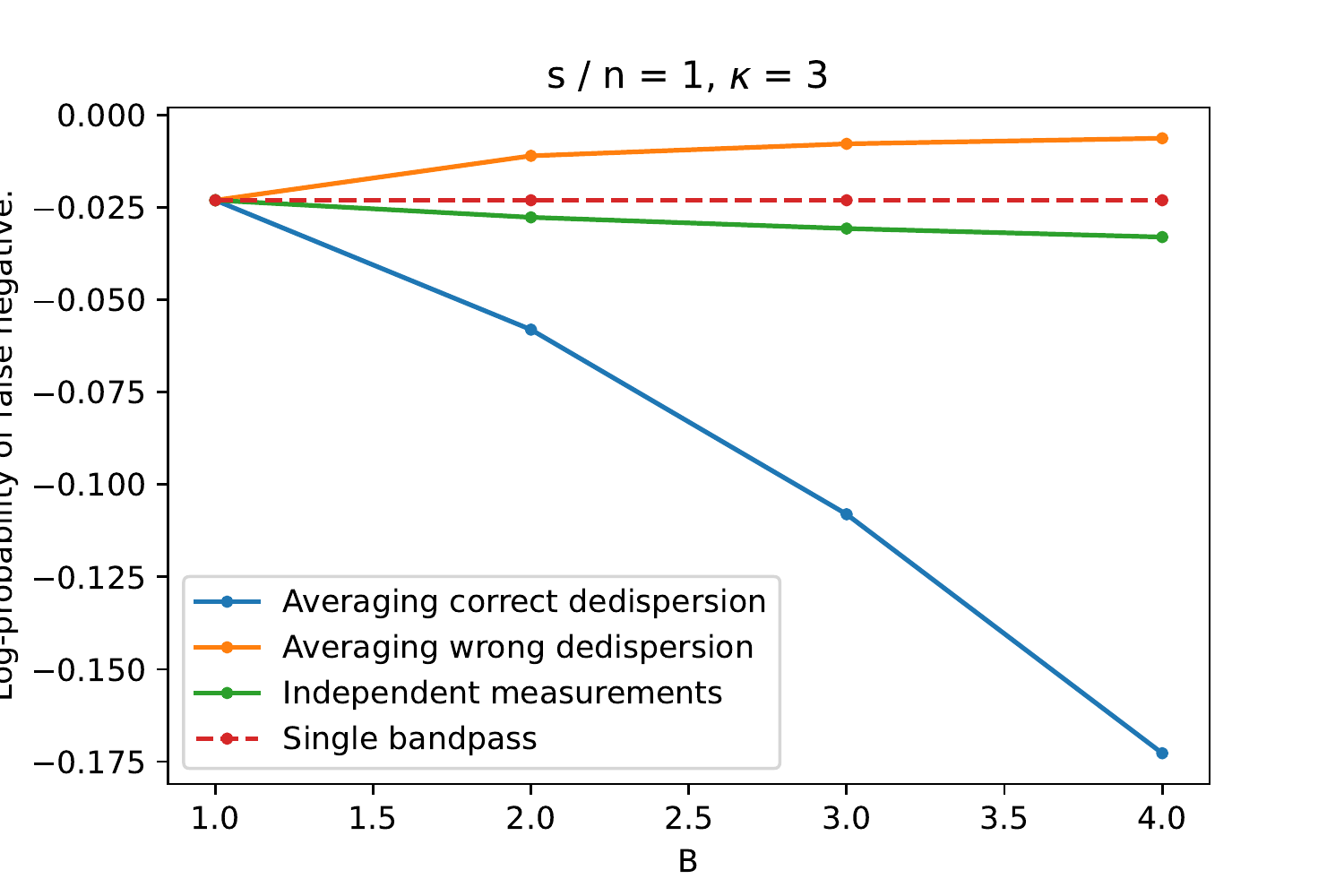}
    \includegraphics[width=0.3\linewidth]{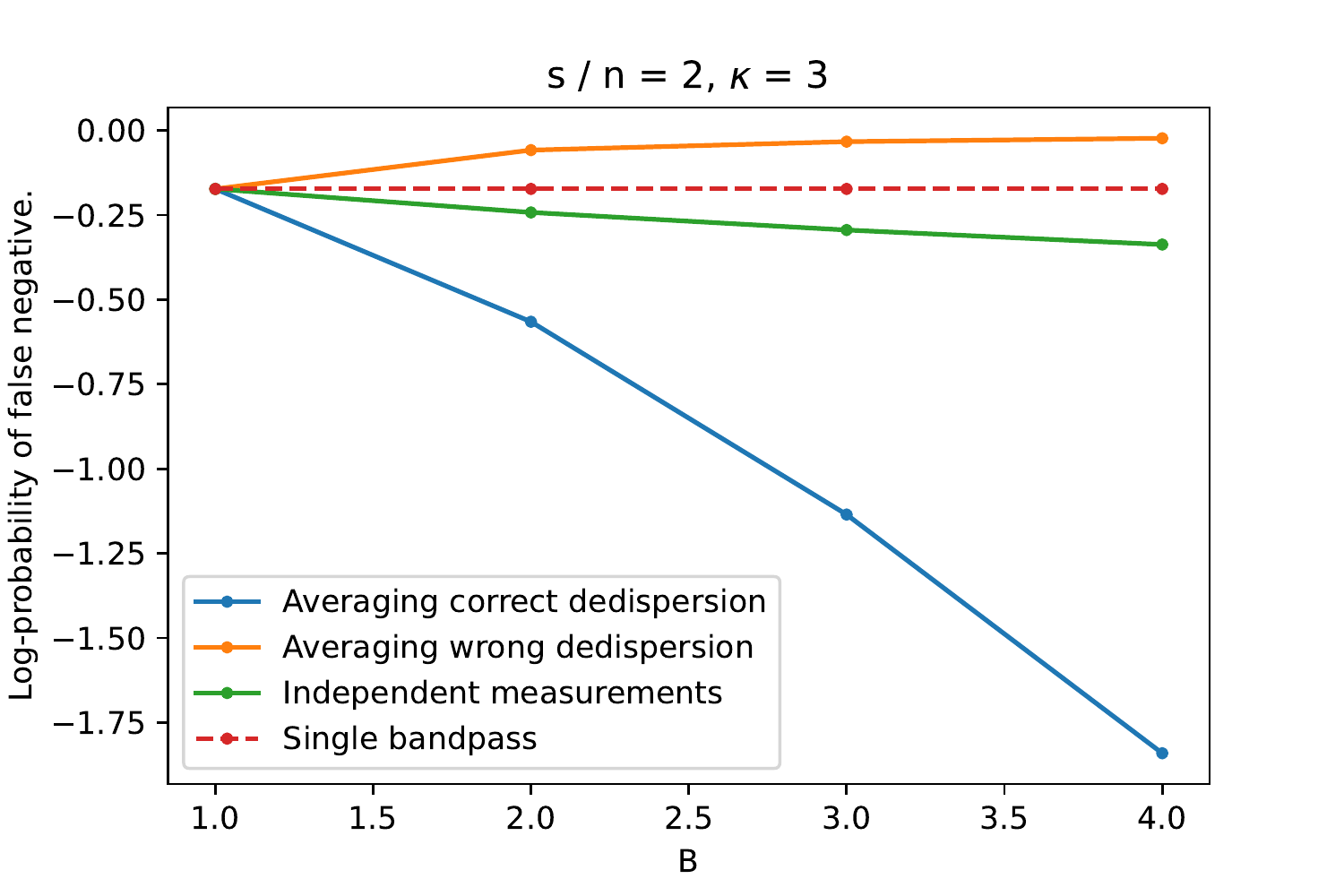}
    \includegraphics[width=0.3\linewidth]{images/appendix/type_2_error_kappa_3_S_5.pdf} \\
    \includegraphics[width=0.3\linewidth]{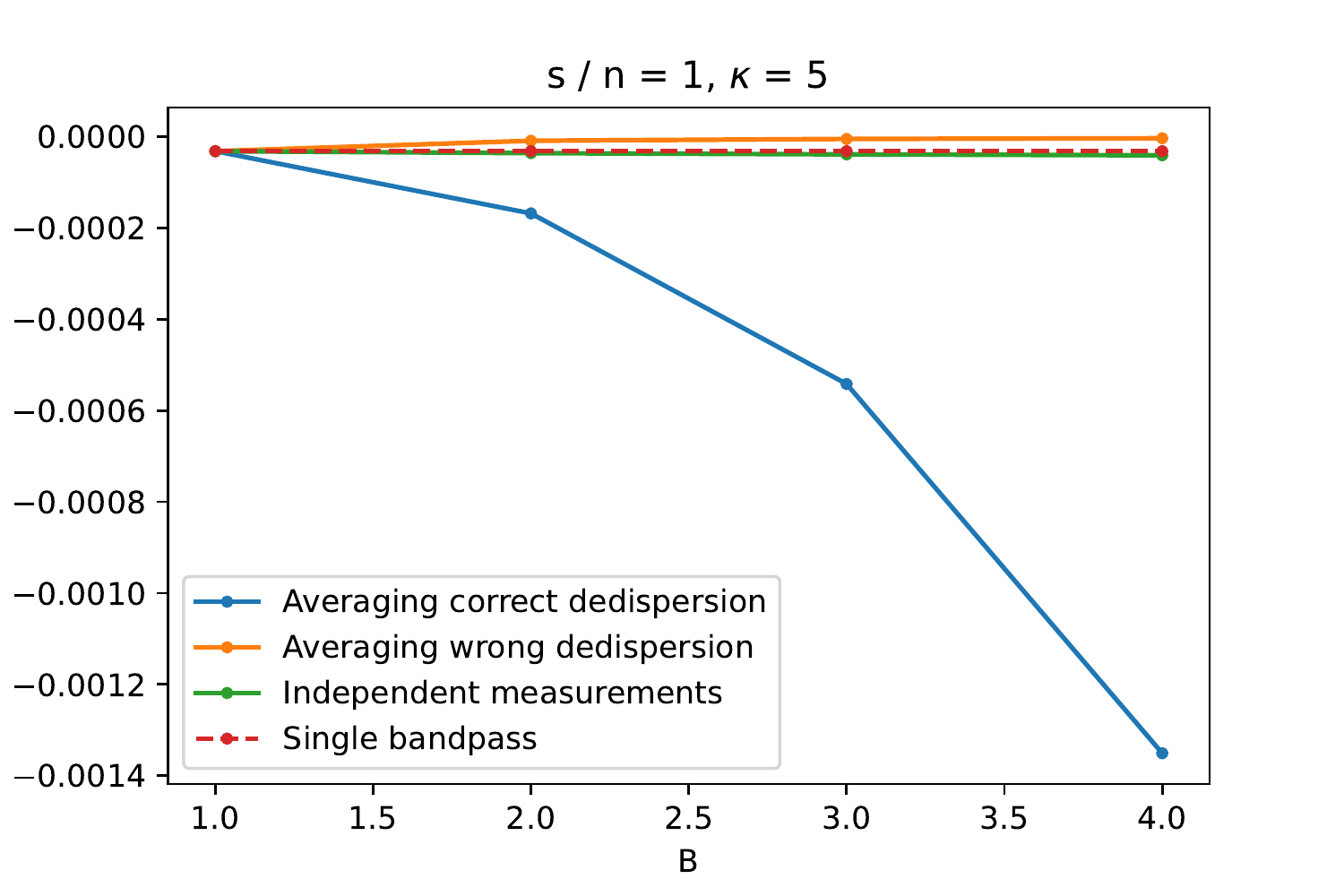}
    \includegraphics[width=0.3\linewidth]{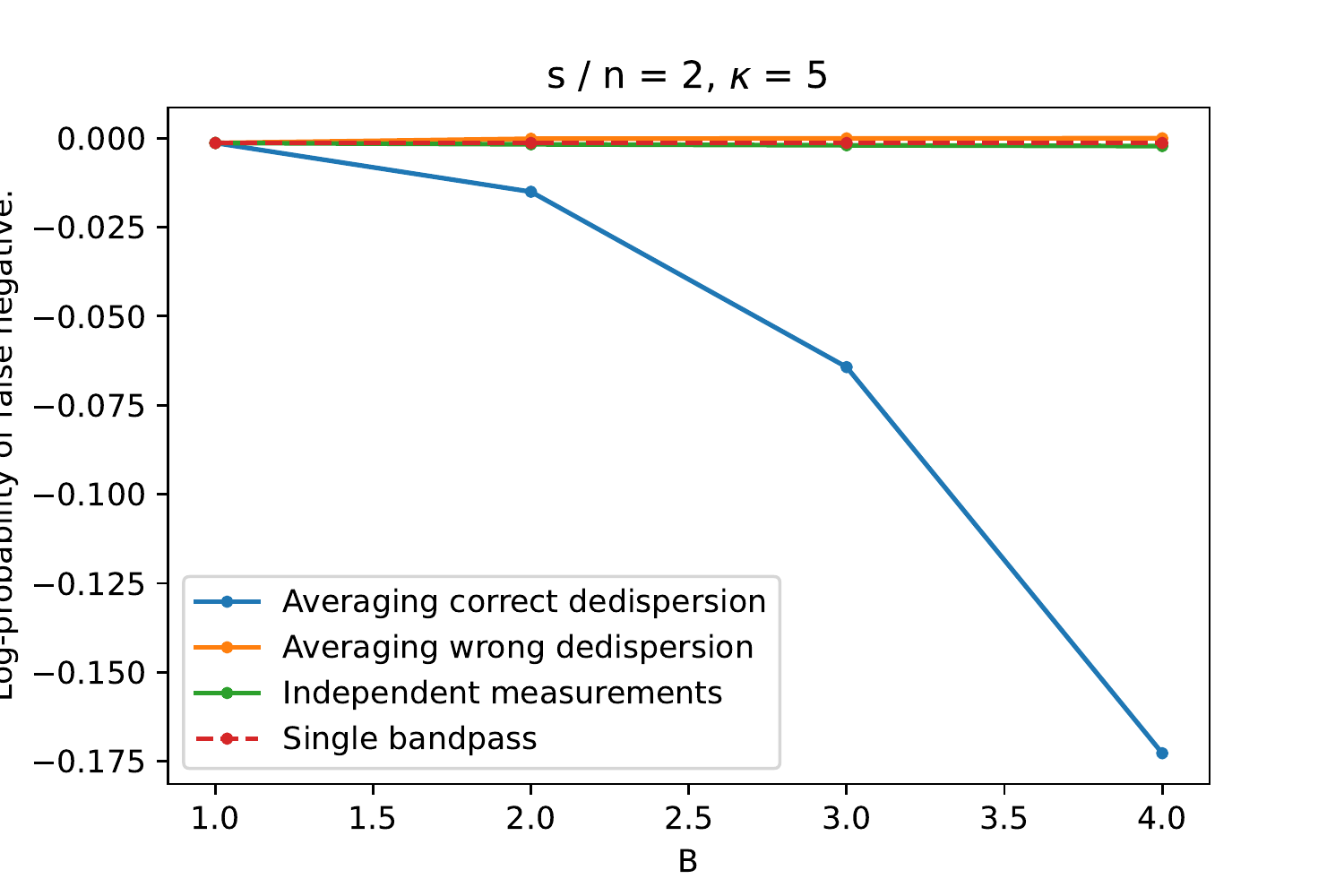}
    \includegraphics[width=0.3\linewidth]{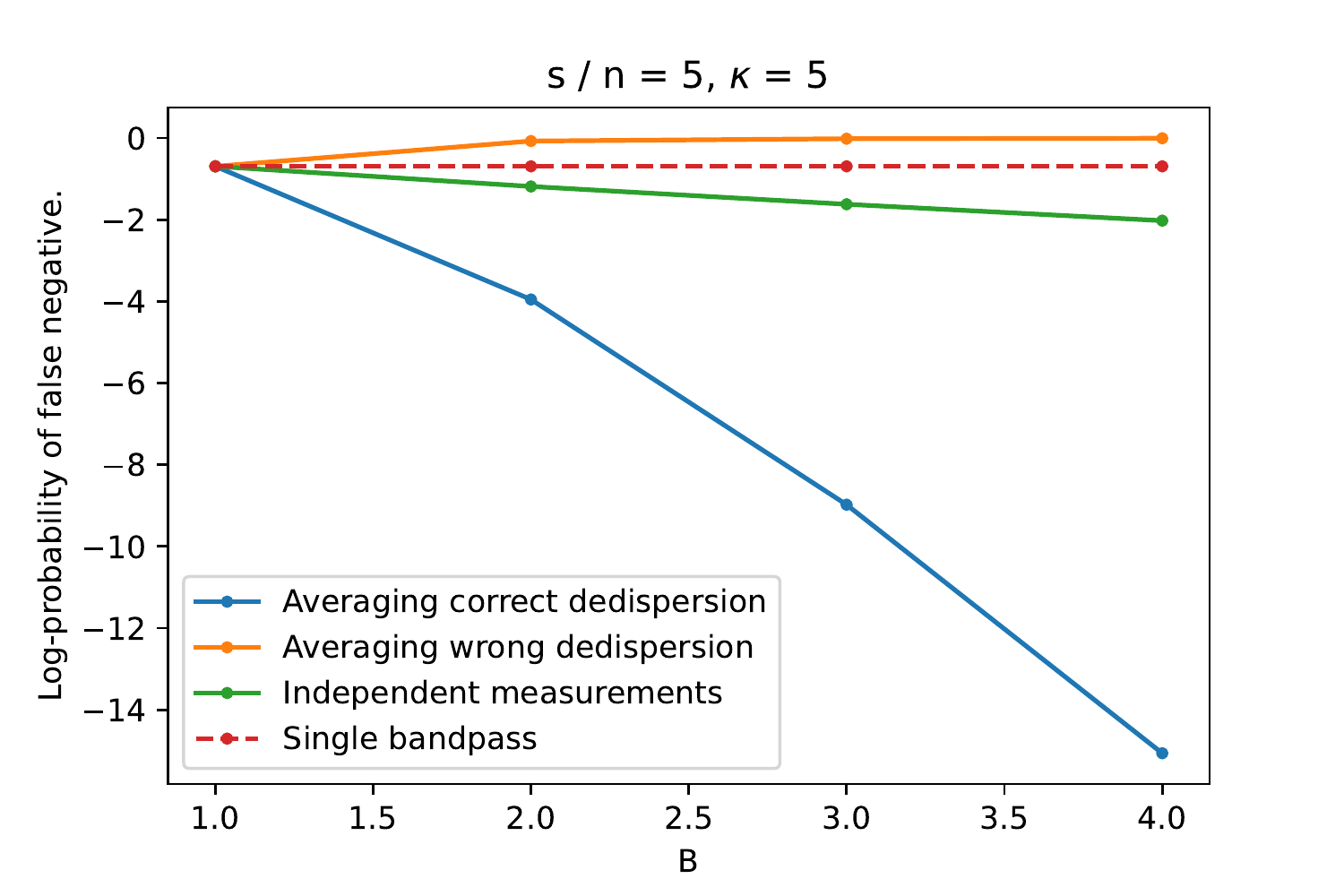} 
    \caption{Results for the statistical comparison of using multi-frequency source detection methods. $y$-axis: log-probability of a false negative (lower is better). $x$-axis: number of passbands $B$.}
    \label{fig:type2_results}
\end{figure*}
\label{app:statistics}
In the following, we perform a statistical analysis on how detecting sources in separate sub-bands individually improves our false negative error rates.
We compare against averaging (with independent noise and constant signal) after applying perfect de-dispersion.
However, note that this is unattainable in practice: one would have to apply source-detection in images de-dispersed against all possible $\operatorname{DM}$. 
This is computationally intractable.
Secondly, the assumptions required are not realistic in practice.
Alternatively, performing detection in the sub-bands individually still gives us a type 2 (false negative) error improvement rate while being computationally feasible.
\subsection{Single Subband}
Consider a signal observed in a single passband. It is corrupted with Gaussian noise:
\begin{equation}
    X := S + E
\end{equation}
with $E \sim \mathcal{N}(0, n^2)$.
We set a threshold that defines our type 1 error rate (i.e., false positives) at $\kappa n$.
Under type 1 error, we consider $S=0$. 
That is, what is the probability that we conclude a detection when there is actually only noise.
The probability of such a type 1 error is 
\begin{equation}
\begin{aligned}
\Pr(X >\kappa n \mid S=0) &= 1 - \Pr(Z \leq \kappa \mid S=0) \\
&=1 - \Phi(\kappa)
\label{eq:type1}
\end{aligned}
\end{equation}
where $Z:= X / n$.
Next, we consider the type 2 error rate.
This expresses the probability that we neglect actual sources ($S = s > 0$), i.e., false negatives.
We have 
\begin{equation}
\begin{aligned}
    \Pr(X \leq \kappa n \mid  S=s) &= \Phi\left(\kappa - \frac{s}{n}\right),
\end{aligned}
\end{equation}
and so our defined $\kappa$ and the signal-to-noise ratio of the source define our probability of neglecting it.

\subsection{Perfect dedispersion}
Consider a signal that we observe in multiple passbands.
We average it after correcting perfectly for the dispersion delay.
Assuming perfect independent Gaussianity with equal variances and equal signals in all channels, our signal-to-noise ratio improves with a factor $B$ (the number of channels).
\begin{equation}
    \bar{X} := \frac{1}{B} \sum_{i=1}^B \left[ S + E_i\right] = S + \bar{E}
\end{equation}

\begin{equation}
\begin{aligned}
    \operatorname{SNR} &:= \frac{S^2}{\mathbb{E}[\bar{E}^2]} \\
    &= \frac{S^2}{\operatorname{Var}[\bar{E}]} \\
    &= \frac{S^2}{\frac1B n^2}
\end{aligned}
\end{equation}

and so the signal-to-noise ratio grows with $\sqrt B$.

Since we keep the false-negative rate fixed, defined by $\kappa n$, our type 2 error improves.
For $S > 0$ we have
\begin{align}
    \Pr(\bar X \leq \kappa n \mid S=s) = \Phi\left(\kappa - \sqrt B  \frac{s}{n}\right). 
\end{align}
With $B > 1$ this is an improvement over the single-band case.

\subsection{Wrong dedispersion}
Often, subbands are averaged in order to improve signal-to-noise ratios.
However, in the face of a dispersed signal, doing so can actually wash out the signal.
In the extreme case, we only observe it in a single band.
Integrating that out would be equal to considering 
\begin{equation}
    \hat X := \frac{1}{B}\left[S + \sum_{i=1}^B E_i \right]
\end{equation}

which is equal to reducing the signal (or boosting the noise) with $\sqrt{B}$.

\begin{equation}
    \operatorname{SNR}:= \frac{\frac{1}{B^2} S^2}{\frac{1}{B} n^2}
\end{equation}

Considering the same type 1 error $\kappa n$ then increases our type 2 error:

\begin{align}
    \Pr(\hat X \leq \kappa n \mid S=s) &= \Phi\left(\kappa - \frac{1}{\sqrt{B}} \frac{s}{n}\right)
\end{align}

\subsection{No dedispersion \& independent detection.}
By doing detection in multiple subbands simultaneously, we are effectively ``throwing the dice'' $B$ times.
Consequently, we obtain an increased type 1 error.
The probability of obtaining a false positive simply increases if you consider more trials.
Thus, to make a fair comparison with the other methods we should adjust for this.

The probability of a type 1 error in this case is (for $S = 0$):
\begin{equation}
\begin{aligned}
&\Pr(Z_1 > \lambda \lor \ldots\lor Z_B > \lambda \mid S=0) \\
&\quad= 1 - \Pr(Z_1 \leq \lambda \land \ldots \land Z_B \leq \lambda \mid S=0) \\
&\quad= 1 - \prod_{i=1}^B \Pr(Z_i \leq \lambda \mid S=0) \\
&\quad= 1 - \left[\Phi(\lambda n)\right]^B
\end{aligned}
\end{equation}
where $\lambda$ now defines the detection threshold and $Z_i := X_i / n$.
To obtain the same type 1 error rate as the other methods (\cref{eq:type1}) we equate them and solve for $\lambda n$.
\begin{equation}
\begin{aligned}
    &1 - \left[\Phi(\lambda n)\right ]^B = 1 - \Phi(\kappa n) \\
    \iff &\Phi(\lambda n) = \left[\Phi(\kappa n)\right]^{1/B} \\
    \iff &\lambda n = \Phi^{-1}(\left[\Phi(\kappa n)\right]^{1 / B})
\end{aligned}
\end{equation}
Thus, we should use $\lambda n$ as a function of $\kappa$ in the current case to make a fair comparison.
Then, the type 2 error ($S = s > 0$) probability is:

\begin{equation}
\begin{aligned}
    \Pr(X_1 \leq \lambda n \land \ldots \land X_B \leq \lambda  n\mid S=s) &= \prod_{i=1}^B \Pr\left(X_i \leq \lambda  n \mid S=s \right) \\
    &= \left[\Phi\left(\lambda  - \frac{s}{n}\right)\right]^B
\end{aligned}
\end{equation}
We inspect how this behaves as a function of $\kappa$ in the following section.

\subsection{Comparison}
Analytically relating these quantities is not trivial due to the CDFs.
Instead, we switch to an empirical study by comparing the type 2 probabilities for different $\kappa, B$, and $S$, fixing $n$ at unity.
The results are shown in \cref{fig:type2_results}.
For small $\kappa$, our method performs relatively well.
This is interesting if running with many type 1 errors (false positives) is not a problem (which is the case for our pipeline).
However,  averaging after dedispersion wins.
The assumptions required usually make this approach unattainable in practice.
\begin{enumerate}
    \item Perfect dedispersion
    \item Uncorrelated noise
    \item Stable signal across all bands
\end{enumerate}
If assumptions 1) and 2) are violated, we may actually risk amplifying the noise and end up in the case where we reduce the signal-to-noise ratio.
Moreover, assumption 1) is infeasible in practice, as we would have to perform source-finding in images averaged for all possible dispersion measures.
However, by running detections in separate subbands we and still gain a type 2 performance increase over not using the multiple bandpasses or wrong dedispersion, which is also frequently done.
Finally, by correcting for dispersion by doing a DM sweep, the probability of a type 1 error increases, which we did not account for in this analysis.
We did take that into account for our method.

\section{Sigma-clip Iterations}
\begin{figure}
    \centering
    \includegraphics[width=0.7\linewidth]{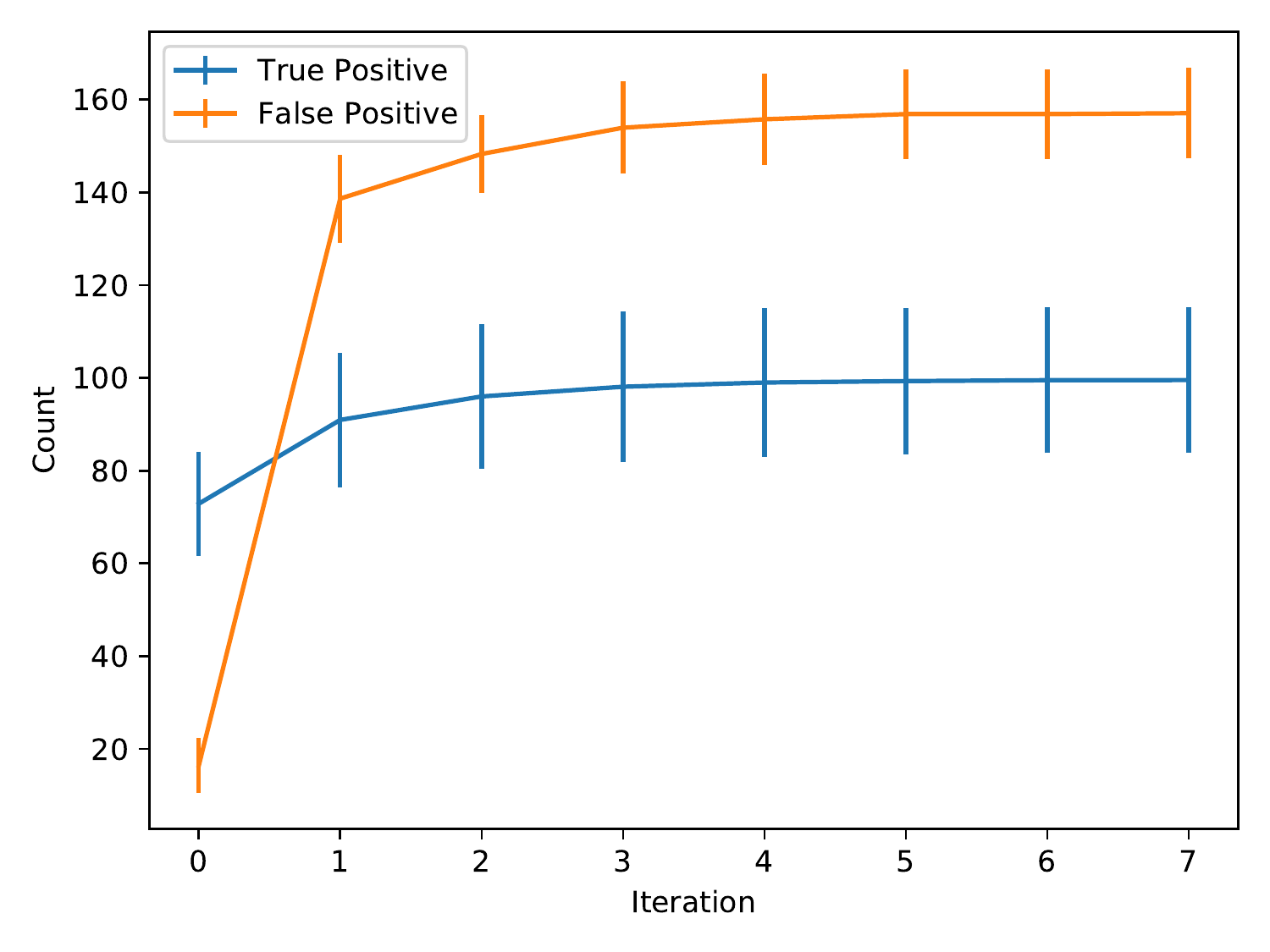}
    \caption{True positive and false positive rate as a function of sigma-clip iterations in our source detection method.}
    \label{fig:iterations_sigmaclip}
\end{figure}
\begin{figure}
    \centering
    \includegraphics[width=0.65\linewidth]{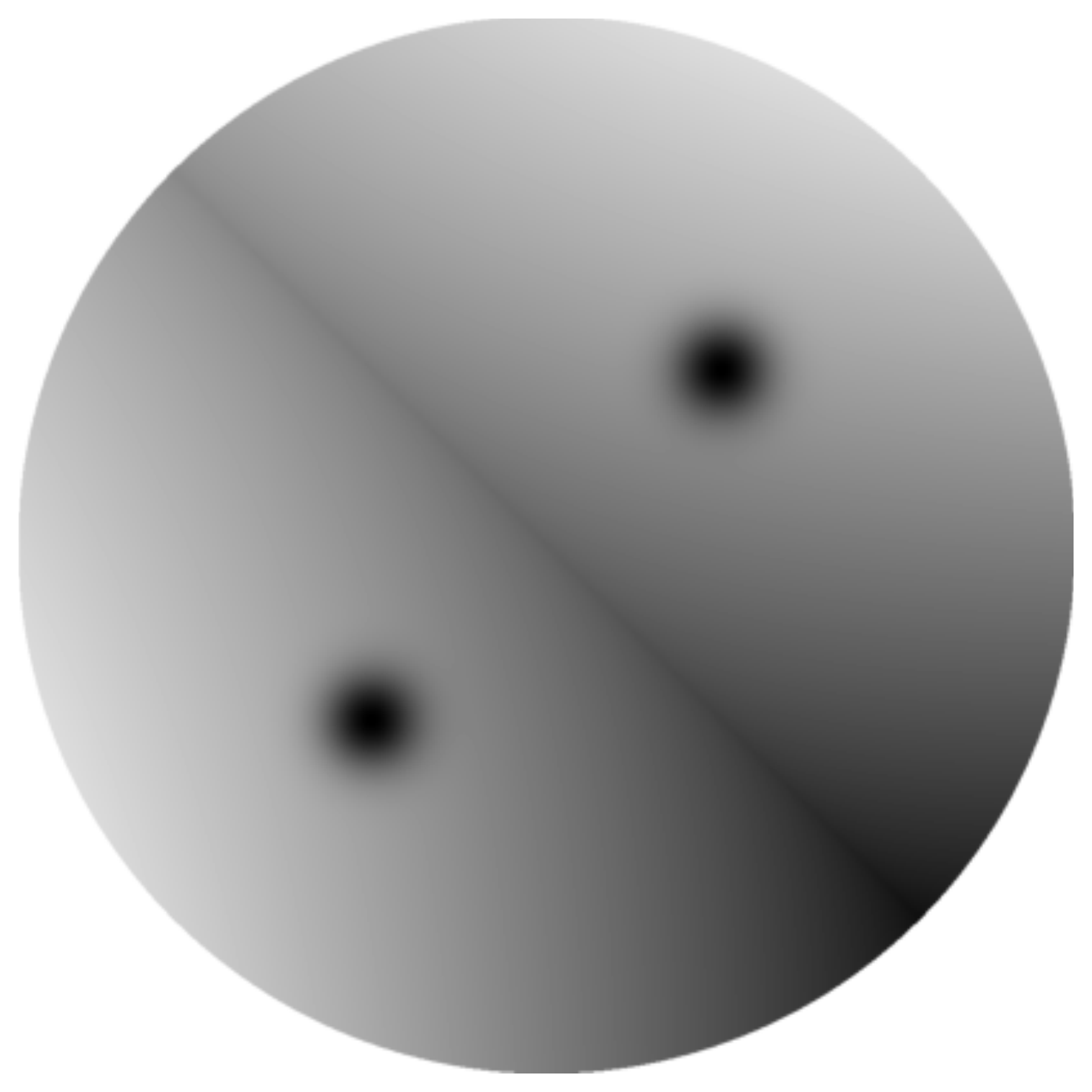}
    \caption{The inhomogeneous galactic foreground \& calibration remnants map used for simulation all-sky images.}
    \label{fig:extended_emission}
\end{figure}
\label{sec:supp_sigmaclip_iterations}
We noted that since we have continuous estimates of the local statistics using convolutions, we do not require many iterations to do peak detection.
This can be seen in \autoref{fig:iterations_sigmaclip}, where the performance of the source-finder is plotted as a function of the sigma-clip iteration.
It is shown that after the first step, we already retrieve many more false positives than true positives.

\section{Skymap Simulation}
\label{sec:supp_sky_simulation}
\begin{figure}
    \centering
    \includegraphics[width=0.65\linewidth]{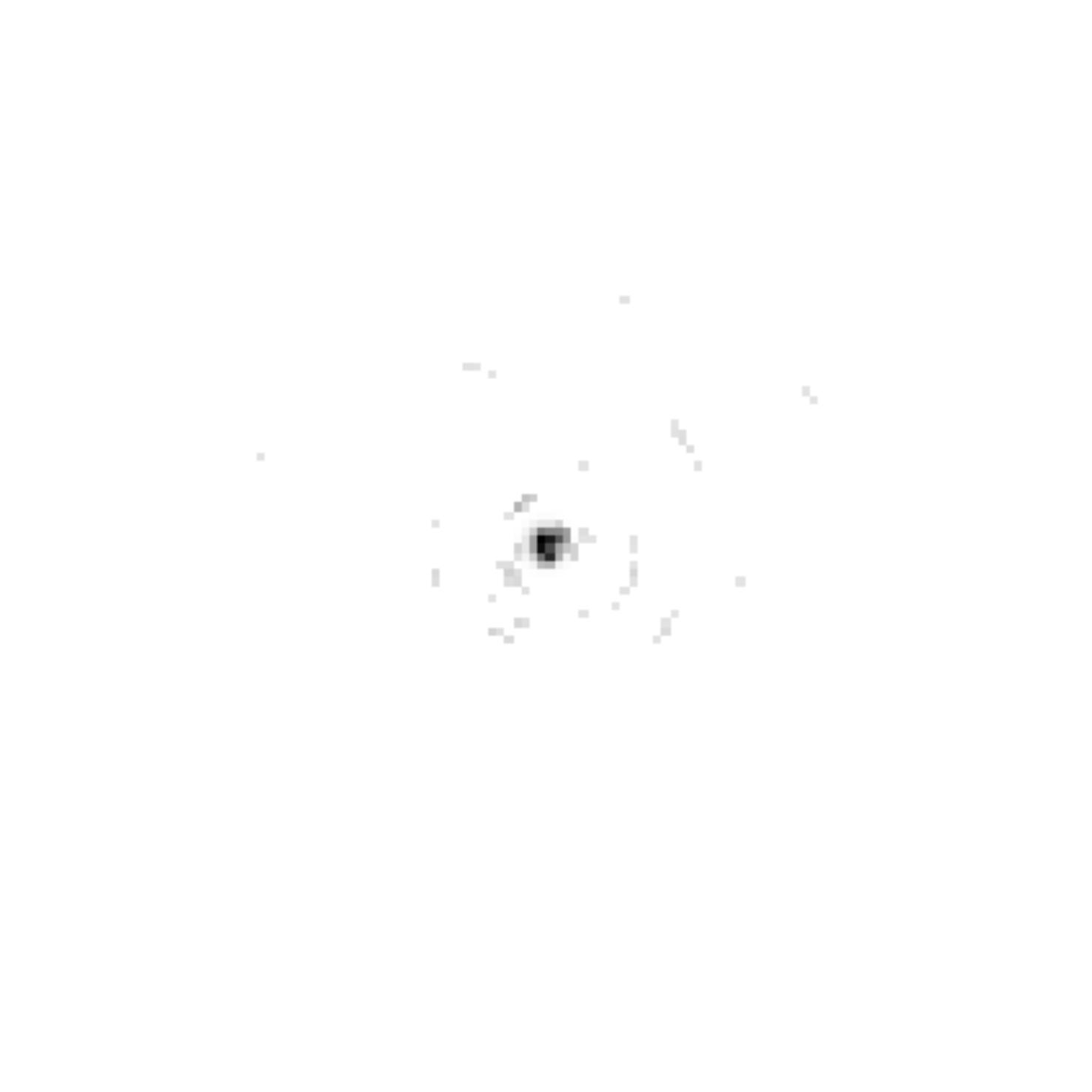}
    \caption{Example of point-spread function that was used to simulate the all-sky images.}
    \label{fig:psf_example}
\end{figure}
\begin{figure}
    \centering
    \includegraphics[width=0.65\linewidth]{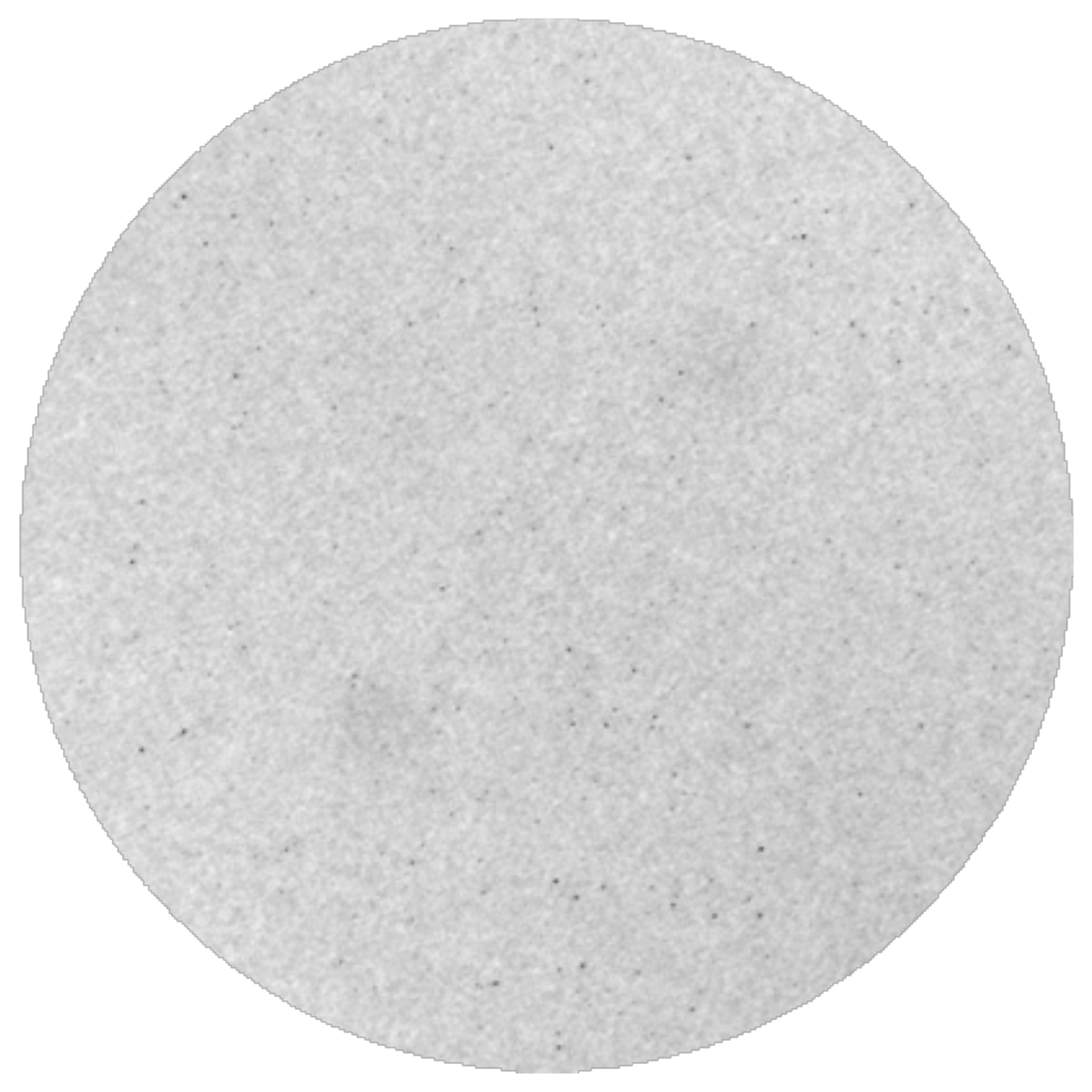}
    \caption{Example of simulated single-channel source map used in our source detection experiments.}
    \label{fig:example_sky}
\end{figure}
In \cref{fig:example_sky} we depict a resulting simulated radio sky.
To test our source-finder we simulate all-sky images and test how many sources we can recover.
Here we detail the procedure.
First, we sample source fluxes from an exponential distribution such that roughly 40\% of the samples have $\mathrm{SNR} < 1$.
This follows \cite{vafaei2019deepsource}.
We sample point source locations uniformly within a circle with diameter $D$.
Noise is sampled from a standard Gaussian.
Additionally, we use a map of inhomogenous extended emission (replicating galactic foreground and calibration remnants) such that the noise levels are not uniformly strong across the entire image (\cref{fig:extended_emission}).
We sample random point-spread functions by augmenting a Gaussian ``backbone'' with random side-lobes that are integrated for 0.2 $\pi$ rad.
An example of such point-spread function is given in \cref{fig:psf_example}.

\end{document}